\documentstyle[eqsecnum, preprint, aps, fixes, floats, epsfig]{revtex}
\tightenlines
\begin{document}

\def\citeNUM#1{b@#1}
\def\citenum#1{\expandafter\csname\citeNUM{#1}\endcsname}
\def\half{{\textstyle {1\over2}}}

\newsavebox{\lsim}
\savebox{\lsim}[1.0em]{\raisebox{-0.3ex}{$\stackrel{<}{\scriptstyle\sim}$}}

\preprint{UW/PT 00-06}
\title{One-Loop Quantum Energy Densities of \\
       Domain Wall Field Configurations}
\author{Andrei Parnachev and Laurence G. Yaffe}
\address{University of Washington, Department of Physics,
Seattle, Washington 98195-1560}
\maketitle

\begin{abstract}
We discuss a simple procedure for computing  
one-loop quantum energies of any static
field configuration that depends non-trivially on only
a single spatial coordinate.
We specifically focus on domain wall-type
field configurations that connect two distinct 
minima of the effective potential, and may or may not
be the solutions of classical field equations.
We avoid the conventional summation of zero-point 
energies, and instead exploit the  relation between
functional determinants and solutions
of associated differential equations.
This approach allows ultraviolet divergences to
be easily isolated and extracted using any convenient 
regularization scheme.
Two examples are considered: two-dimensional 
$\phi^4$ theory, and three-dimensional scalar 
electrodynamics  with spontaneous symmetry 
breaking at the one-loop level.
\end{abstract}

\section{Introduction}

The calculation of the one-loop energy of 
an arbitrary static background field 
configuration is a problem in field theory that has
received much attention over the years.
(For a small sampling of previous papers, see
Refs.~\cite{Dashen&Nevue,SvNR,SRN,Graham&Jaffe,Bashinsky}
and references therein.)
Various complications arise when one
deals with field configurations 
with non-trivial spatial dependence, as extra care 
is required to relate renormalization
of the effective action for such a field configuration
to the renormalization used in the perturbative
sector of a theory.

To illustrate this situation, let us recall
the result for the  kink mass in 1+1-dimensional
$\phi^4$ scalar field  theory \cite{Dashen&Nevue}.
The typical way of computing one-loop corrections
involves the summation over zero-point energies 
computed in the presence of
a kink, and subsequent subtraction of the contribution
from the vacuum sector.
The necessary regularization is a rather
delicate procedure,
and it has been recognized some time ago that the naive 
use of a momentum-cutoff regulator leads to an 
incorrect answer \cite{SvNR,Rajaraman}.
Of course, the fact that the one-loop calculation of
the kink mass could be performed analytically
is directly related to the special form of the 
kink solution for which the spectrum  of the fluctuation
operator in the kink background is exactly known.

For a generic background field configuration,
the situation becomes more problematic
as there is no general procedure
for solving the eigenvalue problem analytically,
and thus one must rely on numerical methods.
This means that the sum over zero-point
energies  must be regularized properly,
and this regularization procedure
must leave one with a well-defined
regulator-independent integral accessible to
numerical evaluation.
The recent paper \cite{Graham&Jaffe} gives one
approach for providing an
unambiguous regularization of the sum over
zero-point energies, based on the subtraction 
of successive Born approximations for scattering
phase shifts.
Having rendered the zero-point sum finite
by this subtraction,
one may then add the subtracted terms back in a 
regularized form, and observe that the regulator dependence
cancels appropriate counterterm contributions contained
in the classical expression for the energy.

However, the approach of Ref. \cite{Graham&Jaffe}
can become complicated as one goes
to higher dimensions (as higher order
approximations for the phase shifts are
needed), or encounter fluctuation operators 
that couple multiple fluctuating fields, as one
inevitably does in gauge theories.
It is therefore desirable to have a method that
would allow a more computationally
straightforward treatment of
ultraviolet divergences.

We will discuss an alternative method, specifically
applicable to kink-type field configurations that 
depend non-trivially on only a single spatial coordinate, which
naturally simplifies the treatment of ultraviolet
divergences and also allows one to deal with
the other problems that appear in the treatment of 
gauge theories.

The crucial feature of the method we describe
is the departure from the conventional
summation over zero-point energies.
Instead, we go back to the functional determinants
that enter the one-loop effective action, and
express them in terms of the
solutions to associated differential equations with
certain boundary conditions.
We examine two specific examples to illustrate
the main ideas.
One natural choice is, of course, the classic
case of 1+1-dimensional scalar  $\phi^4$ field  theory.
This simple case provides us with a consistency
check, as the value of kink mass can
be computed independently by an analytic
calculation.
In addition, working with two-dimensional $\phi^4$
theory presents an opportunity to
illustrate the relevant ideas in a simple
context.

We then examine three-dimensional scalar electrodynamics.
This theory has a first order phase transition 
(as the scalar mass is varied) separating Coulomb
and Higgs phases. 
But the existence of a first order transition, and
hence the presence of a degenerate ground state,
is only apparent when one-loop quantum corrections
are included.
Hence, computing the interface tension (or domain wall
energy density) at the phase transition requires
the evaluation of the effective action of field
configurations which are not classical solutions
of the underlying classical field theory.
To our knowledge, our approach provides the first 
practical technique for computing one-loop
energies of arbitrary Higgs field profiles with
one-dimensional spatial dependence in
a gauge theory.

Three-dimensional scalar QED can be viewed as a low
energy effective theory for four-dimensional scalar
QED at finite temperature.
The domain wall surface tension in the three-dimensional
theory translates into the surface tension of an
interface that separates hot and cold phases of the
finite temperature theory.
This quantity has been a subject of 
several investigations in electroweak theory
\cite{Shaposhnikov,KLRS,Laser}, and scalar
electrodynamics has often been used as a toy
model\cite{Andersen,KKLP}. 
The desire to have a way of correctly 
computing the full one-loop interface tension in
electroweak theory was part of the motivation
for this work.

This paper is organized as follows.
In the next section, we consider two-dimensional
scalar $\phi^4$ theory.
The results of this section are not new, but
this toy model provides an useful playground for
setting up notation, introducing the basic ideas of
our method, identifying possible problems, and developing
appropriate tools for solving them.

In section 3, we  consider three-dimensional
scalar QED.
We show how to deal with the difficulties associated
with the absence of degeneracy of the ground state
at the tree level.
We also show how the scheme used to compute
one-dimensional determinants in section 2 may
be extended to deal with a fluctuation
operator that couples together more than one field.
Numerical results are presented at the end
of this section, where we show that they are
consistent with expected behavior
in the limit of a small ratio of scalar and gauge
couplings.
%
%
%
%
%
%
%
%
%
%
%
%
%
%
%

\section{Two-dimensional  $\mathbf{\phi^4}$ theory}

In this section we will be concerned with our toy model, 
double well scalar $\phi^4$ field theory in two 
dimensions.
We start by formulating
the theory and introducing our notation.
We then review the traditional way of 
calculating one-loop corrections to
the energy of an arbitrary field
configuration\footnote{ In 1+1 dimensions what would be
     the ``energy density'' of a domain wall in higher
     dimensions becomes just the energy, or mass, of
     the kink-like field configuration.}
by summing over zero-point energies.
We implement this prescription and compute 
the one-loop correction to the kink
mass using dimensional regularization.
This warm-up computation, which reproduces old results,
will serve to illustrate the difficulties
that one faces in generalizing this approach
to arbitrary field configurations and more complicated
theories.
We will argue that the integration over one
of the directions (Euclidean time), which reduces 
the functional determinant 
to a sum over zero-point energies, does not
come without a price. 
As we will see, this approach forces
one to resort to computationally inconvenient ways
of regulating ultraviolet divergences.

We then introduce an alternative method that does 
not rely on the summation
of zero-point energies.
We show how one can make the issue of ultraviolet
regularization simple by integrating over the
momentum in the direction of spatial variation
of the field profile, instead of
integrating over the Euclidean frequency.
We relate the resulting one-dimensional functional
determinants to solutions
of associated differential equations, and present
an explicit formula for the mass of an arbitrary
field configuration that connects the two
minima of the effective potential.
The value of the one-loop correction to the kink mass is
reproduced as a check of this general result.


\subsection{One-loop energy density}

Consider scalar  $\phi^4$-theory
with the double well potential
\begin{equation}
\label{phi4potl}
V(\phi) = -\frac{m_0^2}{2} \phi^2+\frac{\lambda_0}{4!}
     \phi^4,
\end{equation}
which exhibits spontaneous symmetry breaking at 
the tree level.
This feature persists quantum-mechanically 
due to the $\phi \rightarrow -\phi$ symmetry.

We consider a Euclidean version of
this theory formulated inside a finite $n$-dimensional
box with Dirichlet boundary conditions for the
fluctuating fields.
Formulating a theory in a variable number of dimensions 
allows us to use dimensional regularization, even
though the final results of this section will only apply to
the two-dimensional theory.

We will be dealing with field configurations
that vary only along a single direction.
We will assign the subscript $\bot$ to this
direction, and the relevant coordinate will be
denoted by $x_{\bot}$.
The coordinates along all other directions are denoted 
by $x_{\|}$.
We will refer to these directions as ``worldvolume''
coordinates in obvious analogy with D-branes in
string theory.

We will denote by $L$ the half-length of
the box in the direction parameterized by
$x_{\bot}$, and by $V_{\|}$ the
volume of an ($n{-}1$)-dimensional slice $x_{\bot}=const$.
Both $L$ and $V_{\|}$ will be taken to
infinity at the end of the calculation, 
leaving us with the theory in infinite Euclidean
space.

We further restrict our attention to 
field configurations $h(x_{\bot})$ that obey the following
boundary conditions:
\begin{eqnarray}
\label{boundary_cond}
   h(-L)&=&-\upsilon, \\
   h(+L)&=&+\upsilon,
\end{eqnarray}
where $\pm \upsilon=\langle \phi \rangle$
are the minima of the effective potential.
The tree-level value of $\upsilon$
is given by
$\upsilon_{\rm tree}^2=\frac{6 m_0^2}{\lambda_0}$.
We require the field profile to approach
the limiting values at $x_{\bot}= \pm \infty$ fast
enough so that the energy density $\sigma$,
which is related to the effective action
$ \Gamma [h] $ as
\begin{equation}
\label{sigma_def}
 \sigma= \lim_{V_{\|}, L \rightarrow \infty}
         \frac{ \Gamma [h] }{V_{\|} },
\end{equation}
is finite.\footnote{We will always normalize
          the effective action so that it
          vanishes for vacuum field 
          configurations: \\ $\Gamma[h \equiv \pm \upsilon]=0$.}

The effective action can be evaluated 
perturbatively and is equal to
the classical action evaluated at the
given field profile plus quantum corrections.
Retaining only one-loop corrections,
one has the following expression for $\sigma$:
\begin{equation}
\label{sigma}
  \sigma  = \int dx_{\bot} \; \left[
    {\cal L}(h(x_{\bot}))- {\cal L}(\upsilon) \right]
    + \lim_{V_{\|} , L \rightarrow \infty} \left[
    \frac{1}{2V_{\|}} 
    \ln \, \det  \frac{\Delta}{ \Delta^{(0)} }
     \right],
\end{equation}
where $\Delta= -\partial_{\bot}^2 -\partial_{\|}^2 + V''$
is the fluctuation operator in the 
non-zero background field sector, while $\Delta^{(0)}$
is the corresponding fluctuation operator in the
vacuum sector.

We can split $\sigma$ into three parts that
correspond, respectively, to the piece of
the classical energy density that contains only
renormalized parameters $m$ and $\lambda$, 
the one-loop quantum corrections,
and the counterterm contribution to the classical
energy density,
\begin{equation}  
\label{splitting}
  \sigma=\sigma_{\rm cl}+\sigma_{\rm qu}+\sigma_{\rm c.t.}.
\end{equation}
The classical piece $\sigma_{\rm cl}$, and the quantum
correction combined with the counterterm
contribution, $\sigma_{\rm qu}+\sigma_{\rm c.t.}$, will
each be UV-finite.

The terms that enter (\ref{splitting}) are
given by
\begin{equation}
\label{sigma_cl}
\sigma_{\rm cl}= \int dx_{\bot} \left[
       \frac{1}{2}(\partial_{\bot} h(x_{\bot}) )^2
     -\frac{m^2}{2} h(x_{\bot})^2+\frac{\lambda}{4!}
     h(x_{\bot})^4 +\frac{3 m^4}{2 \lambda} \right],
\end{equation}
\begin{equation}
\label{sigma_qu1}
 \sigma_{\rm qu}  =  \lim_{L, V_{\|} \rightarrow \infty}
       \left[ \frac{1}{2V_{\|}} 
        {\rm tr} \, \ln \left(
       \frac{ -\partial_{\bot}^2 -\partial_{\|}^2 + V'' }{
          -\partial_{\bot}^2 -\partial_{\|}^2+{V^{(0)}}''}
        \right) \right],
\end{equation}
and
\begin{equation}
\label{sigma_c.t.}
 \sigma_{c.t.}= -\frac{m_0^2-m^2}{2}
           \int dx_{\bot}  \left[ 
            h(x_{\bot})^2-\frac{6 m^2}{\lambda} \right].
\end{equation}

In Eq.~(\ref{sigma_qu1}), $ { V^{(0)} }''=2 m^2$ denotes
the tree-level curvature of the effective potential
at the stationary points.

The fluctuation operators are $x_{\|}$-independent
and can therefore be partially diagonalized by a
Fourier transform over the $x_{\|}$ coordinate.
Taking the $V_{\|} \rightarrow \infty$ limit
recasts $\sigma_{\rm qu}$ as

\begin{equation}
 \label{sigma_qu2}
  \sigma_{\rm qu}= \lim_{L \rightarrow \infty} \left[
        \frac{1}{2} 
           \int \frac{ d^{n-1}k_{\|} }{ (2 \pi)^{n-1} }  
               \ln \, {\rm det}_{\bot} \left( \frac{
		k_{\|}^2-\partial_{\bot}^2+V'' }
		{ k_{\|}^2-\partial_{\bot}^2+ 
		    { V^{(0)} }''           }
	      \right) \right],
\end{equation}
where ${\rm det}_{\bot}$ is a functional determinant
of a linear operator acting on the Hilbert
space  ${\cal H}_L$ of functions
of $x_{\bot}$ that vanish at $\pm L$.


\subsection{Summing over zero point energies}

The conventional approach to computing the one-loop
energy relies on making (\ref{sigma_qu2})
into a sum of zero-point energies.
To accomplish this, one chooses one of the
worldvolume directions to play the role of Euclidean
time, and integrates over the corresponding 
component of $k_{\|}$.
This allows one to rewrite $\sigma_{\rm qu}$ in 
terms of the eigenvalues $ \epsilon_i^2$ of
the operator $-\partial_{\bot}^2+V''$,
\begin{equation}
\label{sigma_qu_sum}
   \sigma_{\rm qu}=  
           \frac{1}{2} 
           \int \frac{ d^{n-2} k_{\|} }{ (2 \pi)^{n-2} }
           \sum_i \left[
             \sqrt{ {\vec k}_{\|}^2+(\epsilon_i)^2 }
             - \sqrt{ {\vec k}_{\|}^2+(\epsilon^{(0)}_i)^2 }
       \right].
\end{equation}
This expression requires the presence of some 
regulator which will render finite both the sum and the 
integration over the remaining worldvolume directions. 
In the two-dimensional case, if we don't use
dimensional regularization, then there are no
integrations left to do and the expression 
for $\sigma_{\rm qu}$
becomes just a suitably
regularized sum over zero-point energies,
\begin{equation}
\label{sigma_qu_2d}
   \sigma_{\rm qu}=  
           \frac{1}{2} 
           \sum_i \left[ \epsilon_i
                 - \epsilon^{(0)}_i \right].
\end{equation}

It is expression (\ref{sigma_qu_2d}) that has 
conventionally been
used for computing the one-loop energies
in (1+1)-dimensional scalar field theories 
\cite{Dashen&Nevue,Graham&Jaffe}.
Below we will re-evaluate the one-loop correction
(\ref{sigma_qu_sum})
to the mass of a kink in the scalar (1+1)-dimensional $\phi^4$
theory using dimensional regularization.
The answer has been known for quite a long
time \cite{Dashen&Nevue}, but it is interesting
to see how dimensional regularization can
naturally avoid various regularization-related subtleties
which appear in more traditional treatments
\cite{Dashen&Nevue,SvNR,SRN}.

The primary purpose of this computation, however,
is to illustrate that the direct use of dimensional
regularization applied to (\ref{sigma_qu_sum})
(without prior subtraction of Born
approximations as in \cite{Graham&Jaffe}) for an
arbitrary field configuration is not desirable.
The computation that we are about to present
is feasible only because the fluctuation operator
for this particular field profile
is nice enough that its spectrum can be 
obtained analytically.

Let us again go back to the theory formulated
in $n$ dimensions, keeping in mind that
physical results will be obtained via 
the limit $n \rightarrow 2$.
The  quartic coupling is only multiplicatively
renormalized to one-loop order:
\begin{equation}
 \lambda_0=\mu^{2-n} \lambda,
\end{equation}
with $\lambda$ bearing the dimension of mass
squared; its beta function vanishes in the physical
limit $n \rightarrow 2$.

The relation between the $\mu$-dependent renormalized
parameter $m(\mu)$ and the bare parameter $m_0$ can be fixed by a calculation of the one-loop  correction
to the self-energy in the
perturbative sector of a theory (which is defined
with respect to one of the degenerate ground states).
This correction has a pole at $n=2$.
To render the self-energy finite, the counterterm
must also have a pole with the same residue.
One has, of course, the freedom to add
some finite quantity to the counterterm. 
We exploit this freedom by choosing
$\overline{MS}$-subtraction, which in
the two-dimensional theory is defined as
\begin{equation}
\label{mass_ren}
  m_0^2=
     m(\mu)^2- \mu^{n-2}\; 
          \frac{\lambda_0}{8 \pi} \left( \frac{
           e^\gamma}{4 \pi} \right)^{n/2-1} 
             \frac{1}{n/2-1},
\end{equation}
where $\gamma$ is Euler's constant.

It is easy to determine the one-loop
beta function of $m(\mu)^2$:
\begin{equation}
  \mu \frac{d}{d \mu} m(\mu)^2= \frac{\lambda}{4 \pi}
   +{\cal O}(\lambda^2).
\end{equation}
To calculate the kink mass we set the
background field configuration $h(x_{\bot})$
equal to the classical kink solution,
\begin{equation}
\label{kink}
   h(x_{\bot})= m_0 \sqrt{6/\lambda_0} \; {\rm tanh}
       \left( \frac{m x_{\bot} }{\sqrt{2}} \right).
\end{equation}

As in (\ref{splitting}), the kink mass 
$\sigma^{\rm (kink)}$ can be split into 
classical, quantum, and counterterm contributions:
\begin{equation}
  \sigma^{\rm (kink)}=\sigma^{\rm (kink)}_{\rm cl}+
       \sigma^{\rm (kink)}_{\rm qu}+
       \sigma^{\rm (kink)}_{\rm c.t.}.
\end{equation}
By direct evaluation of (\ref{sigma_cl}), the
renormalized part of the classical contribution,
which we conventionally denote by $\sigma^{\rm (kink)}_{\rm cl}$,
is given by
\begin{equation}
\label{sigma_cl_exp}
\sigma^{\rm (kink)}_{\rm cl}=4 \sqrt{2} \;
		 \frac{ m(\mu)^3}{\lambda_0},
\end{equation}
while the explicit form of the counterterm
contribution (\ref{sigma_c.t.}) is
\begin{equation}
\label{counterterm}
\sigma_{\rm c.t.}^{\rm (kink)} = 
    -\frac{3 \sqrt{2}}{4 \pi} m \left( \frac{
     e^\gamma}{4 \pi} \right)^{n/2-1} 
     \frac{\mu^{n-2}}{n/2-1}.
\end{equation}

The quantum piece $\sigma^{\rm (kink)}_{\rm qu}$
can be calculated using (\ref{sigma_qu_sum})
and the spectrum of the operator
$-\partial_{\bot}^2+V''$.
Since we are using dimensional regularization, 
we can formally rewrite the sum in (\ref{sigma_qu_sum})
as an integral,
\begin{equation}
\label{kink_mass}
  \sigma^{\rm (kink)}_{\rm qu}=
            \frac{1}{2} \int \frac{ d^{n-2}k_{\|} }{ (2 \pi)^{n-2} }
		 \left[
		   ( k_{\|}^2 )^{1/2} + 
		   ( k_{\|}^2+ \frac{3}{2} m^2 )^{1/2}+
		    \int \frac{ dk_{\bot} }{ (2 \pi) }
		   ( k_{\|}^2 +k_{\bot}^2+ 2 m^2  )^{1/2}
		   \; \delta'(k_{\bot})
		 \right].
\end{equation}
Here $\delta(k_{\bot})$ is the scattering phase shift and
the explicit form  \cite{Dashen&Nevue} of 
$\delta'(k_{\bot}) \equiv \frac{d \delta(k_{\bot})}{dk_{\bot}}$
is
\begin{equation}
\label{delta_prime}
    \delta'(k_{\bot})= -6 \sqrt{2} m \frac{ (k_{\bot}^2+m^2) }{
		(  2 k_{\bot}^2+m^2 ) (k_{\bot}^2+2 m^2 )}.
\end{equation}

The first two terms in (\ref{kink_mass})
come from the two bound states of the small fluctuation
operator in the kink background,
and the last term from the continuous
part of the spectrum \cite{Dashen&Nevue}.
The last term involves a difference between
the contributions from the kink and vacuum
sectors, which do not cancel because of the
differing spectral densities in these sectors. 

Substituting (\ref{delta_prime}) into
(\ref{kink_mass}) one arrives at the following
expression for $\sigma^{\rm (kink)}_{\rm qu}$,

\begin{eqnarray}
\label{sigma_kink_qu}
\nonumber
 \sigma^{\rm (kink)}_{\rm qu}  &=& \displaystyle \frac{1}{2}
   \int \frac{ d^{n-2}k_{\|} }{ (2 \pi)^{n-2} } \Bigg[
     ( k_{\|}^2 )^{1/2} + ( k_{\|}^2+\frac{3}{2}m^2 )^{1/2}
      -3 \sqrt{2} m  \int \frac{d k_{\bot}}{2 \pi} 
      \frac{ m^2 ( k_{\|}^2+ k_{\bot}^2+2 m^2)^{1/2} }{
             (2 k_{\bot}+m^2)(k_{\bot}^2+2 m^2)} 
     \\  \nonumber  && \displaystyle
     -3 \sqrt{2} m  \int \frac{d k_{\bot}}{2 \pi} 
     \frac{k_{\|}^2}{
     ( k_{\|}^2+ k_{\bot}^2+2 m^2)^{1/2} (k_{\bot}^2+2 m^2)} 
     \\    && \displaystyle
     -3 \sqrt{2} m  \int \frac{d k_{\bot}}{2 \pi} 
      \frac{1}{(k_{\|}^2+ k_{\bot}^2+ 2 m^2)^{1/2}} 
\Bigg].
\end{eqnarray}
Let us analyze (\ref{sigma_kink_qu}), term by
term.
The first term that involves $( k_{\|}^2 )^{1/2}$
is set to zero by dimensional regularization,
as it does not contain any external dimensionfull
parameters.
The second, third, and fourth terms are finite, and
equal $3m/2 \sqrt{6}$, $-m/\sqrt{6}$, and
$-3 m /\sqrt{2} \pi$ respectively.
One way to obtain these results is to formally rewrite 
each of the corresponding ($n{-}2$)-dimensional integrals as
a product of an angular integral which is
equal to $2 \pi^{n/2{-}1}/\Gamma(n/2{-}1)$
and a radial integral which can be shown to contain
a $(n/2{-}1)^{-1}$ pole that cancels the zero in inverse 
Gamma function, leading to a finite expression in the 
limit $n \rightarrow 2$.
The last term in (\ref{sigma_kink_qu}) can be 
computed along the same lines,
\begin{equation}
 -\frac{3 \sqrt{2} m}{2}
  \int \frac{ d^{n-2}k_{\|} }{ (2 \pi)^{n-2} } 
  \int \frac{d k_{\bot}}{2 \pi} 
 \frac{1}{(k_{\|}^2+ k_{\bot}^2+ 2 m^2)^{1/2}}=
 -\frac{3 \sqrt{2} m}{4 \pi} 
         \left( \frac{ 2 m^2}{4 \pi \mu^2}\right)^{n/2-1}
         \mu^{n-2} \Gamma(1{-}n/2).
\end{equation}
It obviously produces a pole that cancels the pole in
$\sigma_{\rm c.t.}^{\rm (kink)}$, leaving logarithmic
dependence on the scale $\mu$.
This is completely analogous to what happens
when one does conventional $\overline{MS}$
regularization in the perturbative sector.

We are now in position to collect all the
terms together, and remove the UV regulator by
taking the $n \rightarrow 2$ limit.
This leads to the following expression
for the kink mass:

\begin {equation}
   \sigma^{\rm (kink)}=4 \sqrt{2}\;  
    \frac{m(\mu)^3}{\lambda} + \left[
       \frac{1}{2 \sqrt{6}} 
       -\frac{3}{ \sqrt{2} \pi } 
     + \frac{3 \sqrt{2}}{4 \pi} \,
       \ln{ 2 m^2\over \mu^2 }
     \right] m \,.
\end {equation}
One can check that this expression
is  $\mu$-independent to $ {\cal O}(\lambda)$.

Finally, to make a connection to previous 
results in the literature, we must establish a relation
between the $\overline{MS}$ mass $m(\mu)$ and other
$\mu$-independent physical quantities.
The renormalized mass $M$ used in Ref. \cite{Dashen&Nevue}
was defined by the condition that the 
one-loop effective potential has its minima fixed at 
$\langle \phi \rangle =
  \pm \left( 6 M^2/\lambda \right)^{1/2}$.
This corresponds to the complete cancellation of
the tadpole diagrams by the counterterm.
The relation between $M$ and $m(\mu)$ to one-loop order
is given by
\begin {equation}
\label{renorm_condition}
M^2=m(\mu)^2+ \frac{\lambda}{8 \pi}
\ln \left(2 m^2/\mu^2 \right).
\end {equation}
Substituting this into Eq. (\ref{kink_mass}) gives
\begin {equation}
\label{sigma_kink_exact}
 \sigma^{\rm (kink)}=4 \sqrt{2} \frac{M^3}{\lambda} + \left[
    \frac{1}{2 \sqrt{6}} -\frac{3}{ \sqrt{2} \pi }
     \right] M.
\end {equation}
which coincides with the well-known
result for the kink mass of Ref. \cite{Dashen&Nevue}.

The most important lesson that should be learned
from this calculation is that it would be very
difficult to generalize this dimensional
continuation approach to the case of
an arbitrary field configuration.
Indeed, the discussion below (\ref{sigma_kink_qu})
shows that even after $\sigma_{\rm c.t.}$ cancels the
pole in the quantum part of $\sigma$, taking 
the $n \rightarrow 2$ limit is still a bit subtle.
Namely, it involves the cancellation of a zero
from the angular integral with a pole from the
analytic continuation of the radial integral.
This procedure would be hard to implement in general,
as the radial integral for an arbitrary field configuration
can only be evaluated numerically.
Therefore one must  consider
other ways of regulating a theory, or abandon
the zero-point energy approach.
The former option is chosen in \cite{Graham&Jaffe},
while the latter choice is going to be explored below.


\subsection{Integrating over $k_{\bot}$}

One may compute the one-dimensional
functional determinant in (\ref{sigma_qu2}) directly,
without converting it to a sum over zero
point energies.
The benefit, as we have already advertised, will
be the ability to reduce the computation of the  quantum
correction to the energy of an arbitrary field profile,
$\sigma_{\rm qu}+\sigma_{\rm c.t.}$,
to a finite integral that can be evaluated numerically.

It should be clear that the only reason for 
this procedure not being completely trivial 
is the existence of ultraviolet divergences
in the theory.
Our aim is the establishment of a 
regulator-independent method, which clearly 
separates the divergences and allows one to
use any conventional regularization
scheme in dealing with them.

We will start the analysis of the ultraviolet
divergences by considering an expansion
of the functional determinant in  (\ref{sigma_qu1})
in powers of  
$(\Delta -\Delta^{(0)})/ \Delta^{(0)}$:
\begin{equation}
\label{sigma_qu1_expansion}
 \sigma_{\rm qu}= \lim_{V_{\|} , L \rightarrow \infty} 
    \left[  \frac{1}{2 V_{\|}} 
    \; {\rm tr} \sum_n \; \frac{(-)^n}{n}  
       \left(
         \frac{\Delta-\Delta^{(0)} }{ \Delta^{(0)} }
       \right)^n
     \right].
\end{equation}
In two dimensions, only the first term in this
expansion is divergent.
Recall that  $\Delta^{(0)}$ is just an inverse
bare propagator, so this divergent term can be 
written as
\begin{equation}
\label{divergent_piece2}
 \lim_{ V_{\|},L \rightarrow \infty }
         \left( 
         \frac{1}{2 V_{\|}} 
        \, {\rm tr} \, \left[
          (V''-2 m^2) ( - \nabla^2+2 m^2)^{-1}
          \right] \right)
             =
  \frac{F}{2}  \int \frac{ d^2k }{ (2 \pi)^2 }
         \frac{1}{ k^2+2 m^2 }, 
\end{equation}
where $F$ is the Fourier transform of
$(V''-2 m^2)$ taken at zero momentum,
\begin{equation}
\label{P}
   F=\int dx_{\bot} \left[  (V''-2 m^2) \right]=
   \frac{\lambda}{2}
       \int dx_{\bot} \left[  h(x_{\bot})^2-\frac{6 m^2}{\lambda} \right].
\end{equation}

Of course, the integral in (\ref{divergent_piece2})
is logarithmically divergent.
However, we are free to integrate over 
the momentum in the direction perpendicular
to the worldvolume and then
impose any convenient cutoff 
on the remaining integral.\footnote{
           In contrast, the zero-point energy approach
           does not allow regulator-independent
           integration over $k_{\bot}$,
           as an integration over one of the components
           of $k_{\|}$ has already been performed.
           }
Subtracting this divergent term from (\ref{sigma_qu2}),
and adding it back, we obtain the following
expression for the finite quantity
$\sigma_{\rm qu}+\sigma_{\rm c.t.}$:
\begin{eqnarray}
\label{sigma_qu_22}
\sigma_{\rm qu}+\sigma_{\rm c.t.} & = & \displaystyle
   \frac{1}{2}  \int \frac{ dk_{\|} }{ 2 \pi }
      \left[ 
        \lim_{L \rightarrow \infty}   
        \ln \frac{ \det_{\bot} \left(
		k_{\|}^2-\partial_{\bot}^2+V'' 
		\right) }{
		\det_{\bot} 
		   \left(
		   k_{\|}^2-\partial_{\bot}^2+2 m^2
		  \right)
		}
		- \frac{F}{2 (k_{\|}^2+2 m^2)^{1/2}}
           \right] \\ \nonumber & & \displaystyle
    +\frac{F}{4} \int^{\Lambda} \frac{ dk_{\|} }{ 2 \pi }
     \frac{ 1 }{(k_{\|}^2+2 m^2)^{1/2}}
     +\sigma_{\rm c.t.} \; .
\end{eqnarray}

The first integral in this expression is finite,
as the divergent part has been subtracted from it.
The second term in (\ref{sigma_qu_22}) contains an
integral that appears in the evaluation
of the tadpole diagram, and needs to be regulated.
We indicate this by writing $\Lambda$, which should be
thought of as some cutoff on worldvolume momentum, at
the top of this integral.
However we can use any regulator we like,
as long as it allows unbounded integration over $k_{\bot}$.
The same tadpole diagram also contributes to the
mass renormalization in the perturbative sector,
and consecutively, appears in $\sigma_{\rm c.t.}$.
Comparing (\ref{sigma_c.t.}) and (\ref{P}) 
one can immediately see  that the UV-divergent
terms in (\ref{sigma_qu_22}) cancel each other,
leaving a finite remainder whose value depends on the
renormalization conditions.

Thus, the only non-trivial piece of the calculation
is the evaluation of a single one-dimensional integral.
The expression (\ref{sigma_qu_22}) is, therefore,
a key result of this section.
Note, that had the number of dimensions $n$ been greater
than two, more subtractions might have been needed.
The structure of such terms would be quite clear,
though, as it is determined by the perturbative expansion 
(\ref{sigma_qu1_expansion}).

In what follows we are going to regulate the UV
divergences with dimensional regularization
where the number of worldvolume dimensions
$n{-}1$ is regarded as a variable parameter.
Since the divergent terms in (\ref{sigma_qu_22})
are isolated, the treatment 
that is involved is essentially the same as appeared
when we considered the analytic kink solution.
We choose to express our result in terms
of the physical parameter $M$ defined by
(\ref{renorm_condition}).
One finds simply
\begin{equation}
\label{sigma_qu3}
 \sigma_{\rm qu}+\sigma_{\rm c.t.} =
   \frac{1}{2}  \int \frac{ dk_{\|} }{ 2 \pi } \left[ 
        \lim_{L \rightarrow \infty}   
        \ln \frac{ \det_{\bot} \left(
		k_{\|}^2-\partial_{\bot}^2+V'' 
		\right) }{
		\det_{\bot} 
		   \left(
		   k_{\|}^2-\partial_{\bot}^2+2 M^2
		  \right)
		}
		- \frac{ F }{2(k_{\|}^2+2 M^2)^{1/2}}
           \right].
\end{equation}

When the field configuration is a solution
of the classical equations of motion, the operator
$-\partial_{\bot}^2+V''$ acquires 
a zero eigenvalue, associated with the spontaneous
breaking of translational invariance.
Consecutively, the 2-dimensional fluctuation operator 
$-\partial_{\|}^2-\partial_{\bot}^2+V''$ has continuous
spectra extending down to zero.
In principle one should go back to the original functional 
integral and use standard collective coordinate procedures
\cite{Polyakov} to separate the translational mode.
However, the zero mode of $-\partial_{\bot}^2+V''$ merely
causes the two-dimensional functional determinant
appearing in (\ref{sigma_qu3})
to have an integrable logarithmic singularity 
at $k_{\|}=0$.
Therefore, we do not need to modify  (\ref{sigma_qu3})
for the computation of the one-loop kink mass.

To proceed further, one needs to evaluate
the ratio of one-dimensional functional
determinants appearing in (\ref{sigma_qu3}).
Fortunately, we can use the result of
\cite{Coleman} where this ratio is directly related
to solutions of associated
differential equations. (We review this
construction in  the appendix.)
Specifically, this ratio of one-dimensional determinants
can be written as
\begin{equation}
\label{det1}
    \frac{ \det_{\bot} \left(
		k_{\|}^2-\partial_{\bot}^2+V'' 
		\right) }{
		\det_{\bot} \left(
		   k_{\|}^2-\partial_{\bot}^2+2 M^2 
		   \right)}=
	  \frac{
       \psi(L) }{ \psi^{(0)}(L) },
\end{equation}
where the $\psi$'s are the solutions to the associated
differential equations,
\begin{equation}
\label{deq1}
  \left[ k_{\|}^2-\partial_{\bot}^2+V'' 
 \right] \psi(x_{\bot}) = 0,
\end{equation}
and
\begin{equation}
\label{deq2}
  \left[ k_{\|}^2-\partial_{\bot}^2+2 M^2
 \right] \psi^{(0)} (x_{\bot}) = 0,
\end{equation}
with initial conditions 
\begin{equation}
\label{bc_2}
  \psi(-L)=0; \quad  \psi(-L)'=1,
\end{equation}
and
\begin{equation}
\label{bc_1}
  \psi^{(0)}(-L)  =  0; \quad \psi^{(0)}(-L)'=1.
\end{equation}

The free solution $\psi^{(0)}(x_{\bot})$ is
trivial to find.
It is given by
\begin{equation}
\psi^{(0)}(x_{\bot})=\frac{ \sinh 
          \left[ (k_{\|}^2+2 M^2)^{1/2} (L+x_{\bot}) \right] }{
              (k_{\|}^2+2 M^2)^{1/2} }.
\end{equation}
Neglecting terms which are exponentially small as
$L \rightarrow \infty$, the contribution of this
solution to (\ref{det1}) is
\begin{equation}
\label{psi_vac}
   \psi^{(0)}(L)=\frac{ e^{-(k_{\|}^2+2 M^2)^{1/2} 2 L} }{
             2 (k_{\|}^2+2 M^2)^{1/2} }.
\end{equation}

It is desirable to extract explicitly
the large $L$ asymptotic behavior from $\psi(x_{\bot})$
as well.
This prompts the natural substitution
\begin{equation}
\psi(x_{\bot}) = f(x_{\bot}) 
      e^{-(k_{\|}^2+2 M^2)^{1/2} (x_{\bot}+ L) },
\end{equation}
and converts the ordinary differential equation
(\ref{deq1}) to the corresponding
equation for $f(x_{\bot})$:
\begin{eqnarray}
\label{det2}
\left[ -\partial_{\bot}^2 + 2 (k_{\|}^2+2 M^2)^{1/2}
  \partial_{\bot}+
     \left(V''-2 M^2 \right)
     \right] f(x_{\bot}) = 0,
\end{eqnarray}
with $f(x_{\bot})$ being subject to the same initial
conditions,
\begin{equation}
  f(-L)=0; \qquad f'(-L)=1.
\end{equation}

We can now present the result for the quantum correction
to the mass of an arbitrary kink-like field
configuration in two-dimensional $\phi^4$
theory in its final form,

\begin{equation}
\label{sigma_scalar_final}
 \sigma_{\rm qu}+\sigma_{\rm c.t.} =
   \frac{1}{2}  \int \frac{ dk_{\|} }{ (2 \pi) } \left[ 
        \lim_{L \rightarrow \infty}   
        \ln \left[  2 (k_{\|}^2+2 M^2)^{1/2} f(L) 
            \right]
	- \frac{ F }{2(k_{\|}^2+2 M^2)^{1/2}}
           \right].
\end{equation}

As a check of our method, we have used 
(\ref{det2})--(\ref{sigma_scalar_final})
to recalculate the value of
$\sigma_{\rm qu}+\sigma_{\rm c.t.}$ 
for the kink configuration,
where the exact result is given by
(see Eq. (\ref{sigma_kink_exact}))
\begin{equation}
\sigma_{\rm qu}^{(kink)}+\sigma_{\rm c.t.}^{(kink)}= \left[
    \frac{1}{2 \sqrt{6}} -\frac{3}{ \sqrt{2} \pi }
     \right] M.
\end{equation}
Using the canned routines of
{\it Mathematica} to integrate 
Eq's.~(\ref{det2})--(\ref{sigma_scalar_final}), one may
easily reproduce this result with an accuracy better than 
one part in $10^5$.

%
%
%
%
%
%
%
%
%
%
%
%

\section{Three-dimensional scalar electrodynamics}

We are now ready to tackle the gauge
field theory case.
For the rest of the paper we will be developing
a method for computing one-loop energy densities
(which will sometimes be called the surface or
interface tension) of domain wall-type Higgs
field configurations in three-dimensional scalar 
electrodynamics.
The parameters of this theory will be tuned to sit at
the first order phase transition where, at one-loop level,
there is a degeneracy of the ground state,
and consequently the existence of stable
domain wall-type configurations.

As we will see, the method that we applied
in the previous section can be
successfully generalized to scalar QED.
However, there are two new problems which we did not 
encounter in the scalar theory.
One is the existence of quadratic couplings
between different fluctuating fields in the Lagrangian.
This requires generalization of the result for
a one-dimensional functional determinant.
Another issue is the absence of tree-level degeneracy 
of the effective potential which will cause the quantum and
classical parts of the surface tension to individually
diverge with increasing box size.
We will show how these problems can be resolved. 

%
%
%
%
%
%
%
%
%
%

\subsection{Formulation of the problem}

We will start by formulating the theory and
discussing the relations between the parameters
that are required to tune the theory to its phase 
transition.
We will see that the parameter that determines
the validity of  perturbation 
theory is the ratio of scalar and gauge
couplings.
In the regime where this parameter is small,
it is in fact possible to analytically compute the
leading behavior of the interface tension.
We will later compare the results of this 
analysis with the results of our full 
one-loop computations.

We consider the following Euclidean action for
three-dimensional scalar electrodynamics,
\begin{equation}
\label{qed_action}
 S= \int d^3x  \left[
    \frac{1}{4} F_{ij} F_{ij}+
    \frac{1}{2}(D_i \Phi)^{\dagger}(D_i \Phi)+V(\Phi)+
    {\cal L}_{\rm g.f.}+{\cal L}_{\rm gh}
          +{\cal L}_{\rm c.t.}
	  \right],
\end{equation}
where $D_i \equiv \partial_i+i g A_i$ denotes the usual covariant derivative
with respect to the abelian gauge field $A_i$, and
$ {\cal L}_{\rm g.f.}$ and ${\cal L}_{\rm gh}$ 
denote gauge fixing and ghost 
contributions to the Lagrangian, respectively, which
will be specified later.
We will eventually be using dimensional regularization,
so the counterterm Lagrangian ${\cal L}_{\rm c.t.}$
would in general contain $(n{-}3)^{-1}$ poles
where $n$ is the analytically continued
number of dimensions.
In three-dimensional scalar QED, however, no poles appear
at one-loop order.
Hence, we may drop explicit mention of counterterms 
in all subsequent calculations.

We choose the simple quartic potential for 
the complex Higgs field $\Phi$,
\begin{equation}
  V(\Phi)=\frac{m^2}{2}|\Phi|^2 + 
          \frac{\lambda}{4!} |\Phi|^4,
\end{equation}
with a positive coefficient $m^2$ in front of the
quadratic term.

Let us briefly sketch the relations between the
parameters of the theory that follow from the
condition of a doubly degenerate ground state.
Recall that because of the $U(1)$ gauge symmetry,
the effective potential $V_{\rm eff}$ may be chosen
to be a function of a real expectation value of
the Higgs field.
The crucial observation is that for the theory
at hand, the effective potential  acquires a
$(-g^3 \Phi^3)$ term which comes from the summation
of all one-loop gauge field diagrams with an arbitrary
number of insertions of $g^2 \Phi^2$ \cite{Arnold&Espinosa}.
The existence of two degenerate minima of the
effective potential hence implies that 
the effective potential acquires the form of a double
well,
\begin{equation}
\label{veff_double_well}
V_{\rm eff} \approx \frac{m^2}{2 \upsilon^2}
           \Phi^2 (\Phi-\upsilon)^2,
\end{equation}
where $\upsilon$ is the expectation value
of $\Phi$ in the Higgs phase.
This immediately implies the following characteristic
relations between the parameters:
\begin{eqnarray}
\label{relations_atv}
  \upsilon \sim  g^3/\lambda, \\ \nonumber
 m^2 \sim g^6/\lambda.
\end{eqnarray}

Given (\ref{relations_atv}), one
can verify that the parameter controlling the reliability
of perturbation theory
in both the symmetric and Higgs phases is
$\lambda/g^2$ \cite{Arnold&Espinosa}.
For example, adding an extra loop in a 
diagram in the Higgs phase due to
the addition of a gauge field propagator
leads to an additional  factor of order
$g^2 \int d^3k (k^2+M_W^2)^{-2} \sim g^2/M_W \sim \lambda/g^2$, 
where $M_W=g \upsilon$ is the mass of a gauge
boson.
Thus, perturbation theory is reliable for
small $\lambda/g^2$, even though the existence
of a Higgs phase is not apparent at tree level,
and is caused only by quantum corrections.

As in the previous section, we put the system into a three-dimensional
box of half-length $L$ and transverse area $V_{\|}$.
The computation of the
one-loop surface tension amounts to the evaluation
of the one-loop effective action for a given field
profile $h \equiv h(x_{\bot})$, which we may choose to
be real and positive.
We will restrict our attention to 
domain wall-type configurations that obey 
boundary conditions similar to (\ref{boundary_cond}):
\begin{eqnarray}
 h(-L)=0, \\
 h(L)=\upsilon.
\end{eqnarray}
We decompose the complex Higgs field $\Phi$ 
into the background field $h$ plus two real fluctuating fields,
\begin{equation}
   \Phi=h+\chi+i \phi,
\end{equation}
and attribute gauge variations to the fluctuating
fields $\chi$ and $\phi$.

The effective action is a sum of
classical and one-loop terms:
\begin{equation}
\label{gamma_splitting}
 \Gamma[h]=\Gamma_{\rm cl}[h]+
              \Gamma_{\rm qu}[h],
\end{equation}
where
\begin{equation}
\label{gamma_cl}
\Gamma_{\rm cl}[h]=V_{\|} \int dx_{\bot} \left[
  \frac{1}{2} (\partial_{\bot} h(x_{\bot}) )^2+
      \frac{m^2}{2}  h(x_{\bot})^2+ 
        \frac{\lambda}{4!} h(x_{\bot})^4
          \right].
\end{equation}

To get the quantum part, we need to expand the
action (\ref{qed_action}) around the background
field configuration, keeping terms that
are quadratic in fluctuating fields.
In doing so, one notices that a term of
the form $A_i \partial_i \phi$ appears.
In the presence of a spatially varying background
field, it is not possible to eliminate entirely
this cross-coupling term.
But one may partly eliminate
this term by choosing a modification
of $R_{\xi}$ gauge with the following 
gauge-fixing term,
\begin{equation}
\label{gauge_fix}
   {\cal L}_{\rm g.f.}=\frac{1}{2 \xi}
     \left[ \partial_i A_i + \xi \; \frac{g}{2} \;
                           {\rm Im} (\Phi^2) \right]^2.
\end{equation}
The part of the action quadratic in
fluctuating fields then reads
\begin{equation}
  S^{(2)}=\int d^3x \left( {\cal L}_{\chi}+
       {\cal L}_{A}+ {\cal L}_{\phi}
       +{\cal L}_{A \phi}+{\cal L}_{\rm gh} \right),
\end{equation}
where
\begin{eqnarray}
\label{L_chi}
{\cal L}_{\chi} & = &\frac{1}{2} (\partial_i \chi)^2+
       \left[ \frac{m^2}{2}+\frac{\lambda}{4} 
                     h(x_{\bot})^2 \right]
                   \chi^2, \\
{\cal L}_{A} & = &\frac{1}{2} A_i \left( -\delta_{ij}
                 \partial^2-(1-\xi^{-1})\partial_i
                                              \partial_j
                +g^2 h(x_{\bot})^2 \delta_{ij}
                         \right) A_j \; , \\
{\cal L}_{\phi} & = &
     \frac{1}{2} (\partial_i \phi)^2+
         \left( \frac{m^2}{2}+\left[
                \frac{\lambda}{12}+\frac{\xi}{2}g^2 \right]
                         h(x_{\bot})^2  \right) \phi^2, \\
{\cal L}_{A \phi} & = & -2 g (\partial_{\bot} h(x_{\bot}))
                        A_{\bot} \phi,\\
\label{L_gh}
{\cal L}_{\rm gh} & = & \bar{\eta} \left( -\partial^2 +
                         \xi g^2 h(x_{\bot})^2 \right)
                                \eta.
\end{eqnarray}

It is particularly convenient to choose $\xi=1$, 
because with this gauge choice the terms containing
$(\nabla A)^2$ disappear.
Notice that only the $A_{\bot}$ component
of the gauge field is coupled to the $\phi$ field,
thanks to the gauge choice (\ref{gauge_fix}).

Performing the functional integral 
gives a product of functional determinants, as always.
One can verify that with the gauge choice that we have made,
the contribution of the ghosts completely
cancels the contribution of the two $A_{\|}$ components 
of the gauge field.

The expression for $\Gamma_{\rm qu}$ is thus composed
out of contributions from the Higgs boson $\chi$
and the coupled fields $A_{\bot}$ and $\phi$:

\begin{eqnarray}
\label{gamma_qu}
\displaystyle
 \Gamma_{\rm qu}[h]
    & = & \Gamma_{\chi}[h]+\Gamma_{A\phi}[h], \\
\label{gamma_chi}
\displaystyle
\Gamma_{\chi}[h] &=& \ln \, \det  
          \left( \frac{\Delta_{\chi}}{ \Delta^{(0)}_{\chi} }
   \right), \\
\label{gamma_aphi}
\Gamma_{A\phi}[h] &=& \ln \, \det \left(
          \frac{\Delta_{A\phi}}{ \Delta^{(0)}_{A\phi} }
  \right),
\end{eqnarray}
where the fluctuation operators can be read off
directly from (\ref{L_chi})--(\ref{L_gh}).
Before we write them explicitly, let us introduce
notation that will simplify the
subsequent formulae, by defining
\begin{eqnarray}
\label{m_h}
M_H(h)^2&\equiv&m^2+\frac{\lambda}{2} h^2, \\
M_W(h)^2&\equiv&g^2 h^2, \\ 
M_{A\phi}(h)^2&\equiv& m^2+(\frac{\lambda}{6}+g^2) h^2.
\end{eqnarray}
With this notation, the operator that determines
quadratic fluctuation of $\chi$ is
\begin{equation}
\label{Delta_chi}
 \Delta_{\chi}= -\nabla^2 + M_H(h)^2,
\end{equation}
where according to (\ref{m_h}),
$M_H(h)^2=m^2+\frac{\lambda}{2} h(x_{\bot})^2$.
The fluctuation operator that couples $A_{\bot}$ 
and $\phi$ is denoted by $\Delta_{A\phi}$.
It is a linear operator on 
${\cal H}_L \bigoplus {\cal H}_L$,
and can be represented by the $2 \times 2$ matrix:
\begin{equation}
\label{Delta_aphi}
\Delta_{A\phi}=
\left( \begin{array}{ccc}
   -\nabla^2 + M_W(h)^2 & -2 g (\partial_{\bot} h(x_{\bot}) )\\
   -2 g (\partial_{\bot} h(x_{\bot})) &  -\nabla^2 + M_{A\phi}(h)^2
\end{array} \right).
\end{equation}
In Eq.~(\ref{gamma_aphi}), $\Delta^{(0)}_{A\phi}$
denotes the corresponding
operator in the symmetric vacuum sector.\footnote{
   We could equally well have 
   chosen $\Delta^{(0)}_{A\phi}$
   to be the fluctuation operator in the Higgs phase,
   since the two ground states are degenerate}

We can rewrite the expressions (\ref{gamma_chi})
and (\ref{gamma_aphi}) using the fact that the
dependence on worldvolume coordinates of these 
operators is trivial, and thus they can be partially
diagonalized by a Fourier transform.\footnote{
      This is what we also did in section 2, by
      rewriting $\sigma_{\rm qu}$ as an integral over
      the worldvolume momentum.}
That is,
\begin{eqnarray}
\label{gamma_chi_intd}
\Gamma_{\chi}[h]=\frac{V_{\|}}{2} \int 
       \frac{d k^{n-1}_{\|}}{ (2 \pi)^{n-1} } \;
       \ln \; \frac{ \det_{\bot} \tilde{\Delta}_{\chi} }{
                   \det_{\bot} \tilde{\Delta}_{\chi}^{(0)}},  \\
\label{gamma_aphi_intd}
\Gamma_{A \phi}[h]=\frac{V_{\|}}{2} \int 
       \frac{d k^{n-1}_{\|}}{ (2 \pi)^{n-1} } \;
       \ln \; \frac{ \det_{\bot} \tilde{\Delta}_{A \phi} }{
                    \det_{\bot} \tilde{\Delta}_{A \phi}^{(0)}},
\end{eqnarray}
where
\begin{equation}
  \tilde{\Delta}_{\chi}=k_{\|}^2-\partial_{\bot}^2+M_H(h)^2,
\end{equation}
and
\begin{equation}
\tilde{\Delta}_{A \phi}=
\left( \begin{array}{ccc}
   -\partial_{\bot}^2 + k_{\|}^2+M_W(h)^2 & -2 g (\partial_{\bot} h(x_{\bot}))\\
   -2 g (\partial_{\bot} h(x_{\bot})) &  -\partial_{\bot}^2 +  k_{\|}^2+M_{A\phi}(h)^2
\end{array} \right).
\end{equation}

As we saw in the previous section, the 
divergent integrals over the worldvolume momentum in
(\ref{gamma_chi_intd}) and (\ref{gamma_aphi_intd}) 
can be regularized using any convenient regulator. 
We just need to subtract the divergent pieces 
and add them back in a regularized form.
We find it convenient to use dimensional regularization.

One can compute the effective potential
by taking the field configuration to be
coordinate-independent and going to the limit
$L, V_{\|} \rightarrow \infty$.
Then the operators  $\tilde{\Delta}_{\chi}$
and  $\tilde{\Delta}_{A \phi}$ can be further
diagonalized, giving the following expression
for the effective potential:
\begin{eqnarray}
\nonumber
 V_{\rm eff}(\Phi) &=&  \displaystyle
 \frac{m^2}{2} \Phi^2 {+} \frac{\lambda}{4!} \Phi^4
 \\ && \displaystyle
  {+} \frac{1}{2} \int \frac{d^n k}{(2 \pi)^2} \left[
    \ln \left( \frac{k^2{+}M_H(\Phi)^2}{k^2{+}m^2} \right)
  {+} \ln \left( \frac{k^2{+}M_W(\Phi)^2}{k^2} \right)
  {+} \ln \left( \frac{k^2{+}M_{A\phi}(\Phi)^2}{k^2{+}m^2} \right)
  \right].
\end{eqnarray}

The integrals can be easily evaluated using
dimensional regularization.
The resulting effective potential 
(which has also been computed in many 
references, see {\it e.g.} Refs. \cite{Shaposhnikov,Andersen,KKLP})
reads
\begin{equation}
\label{effective_potential}
\displaystyle
 V_{\rm eff}(\Phi) =  \displaystyle
 \frac{m^2}{2} \Phi^2+ \frac{\lambda}{4!} \Phi^4-\frac{1}{12 \pi}\left(
  M_H(\Phi)^3+M_W(\Phi)^3+M_{A\phi}(\Phi)^3-2 m^3\right) .
\end{equation}

As we asserted earlier, a cubic
$g^3 \Phi^3=M_W(\Phi)^3$ term has
appeared, leading to the relations (\ref{relations_atv})
for the parameters of the theory.

%
%
%
%
%
%
%
%
%
%

\subsection{Computation of the surface tension}

The energy density of a domain wall-type
configuration is a sum of the classical and 
quantum parts, as defined in  (\ref{gamma_splitting}).
Both classical and quantum terms immediately
follow from (\ref{gamma_cl}) and (\ref{gamma_qu}),
by dividing the corresponding expressions by the
transverse area $V_{\|}$.
Note, that $\sigma_{\rm cl}$ is now linearly 
divergent with increasing $L$, as there is no ground
state degeneracy in the tree-level effective
potential.
This means that the quantum part of
$\sigma$ must diverge with increasing $L$ as well,
so that the total $\sigma$ is finite as 
$L \rightarrow \infty$.
Hence, we must take the $L \rightarrow \infty$ 
limit at the end of the calculation, after
all terms are collected together.

The quantum part, $\sigma_{\rm qu}$, can be further
decomposed analogously to (\ref{gamma_qu}) with
corresponding terms called $\sigma_{\chi}$ and
$\sigma_{A \phi}$, respectively.

Let us concentrate on the calculation of  $\sigma_{\chi}$
first.
We would like to apply the same technique for
computing the ratio of determinants as used
in the previous section.
However the situation is now different,
since $\displaystyle \upsilon=
\lim_{x_{\bot} \rightarrow \infty} h(x_{\bot})$
is no longer the minimum of the tree-level
effective potential, and thus the ratio
of determinants diverges with $L$ instead of
being asymptotically constant.
In fact, it must diverge exponentially, so
that after taking the logarithm, the leading
divergence is linear in $L$.

Naturally, we would like to extract this
asymptotic behavior, leaving to the computer
only a calculation of a term that is finite in the limit 
$L \rightarrow \infty$. 
For that purpose, we impose a restriction
on our field configuration.
We require it to be constant outside the region
$-a \le x_{\bot} \le a$; so that
$h(x_{\bot})=0$ for $x_{\bot}<-a$ and
$h(x_{\bot})=\upsilon$ for $x_{\bot}>a$.
In other words, we require the domain wall field
profile to have a ``thickness'' no larger than $2a$.
Eventually, $a$ will be taken sufficiently large
so that this condition has negligible effect.

Thus there now are two length parameters 
that we have introduced:
$a$, being the half-length of the window where
the background field is allowed to vary,
and $L$, the half-length of the box in the
direction perpendicular to the worldvolume.
The quantum part of the surface tension,
as well as $\sigma_{\chi}$ and $\sigma_{A \phi}$
individually, are going to be asymptotically constant
in $a$, but linearly divergent in $L$ as we 
send $L$ to infinity.

The ratio of determinants in $\sigma_{\chi}$ 
can be written in terms of the solution of
associated ordinary differential equations, just
as in Eq.~(\ref{det1}).
This leads to the following expression for 
$\sigma_{\chi}$:
\begin{equation}
\label{sigma_chi}
\sigma_{\chi} =  \frac{1}{2} 
    \int \frac{ d^{n-1}k_{\|} }{ (2 \pi)^{n-1} } 
   \ln \left( \frac{\psi(L)}{\psi^{(0)}(L)} \right),
\end{equation}
where $\psi(x_{\bot})$ is the solution 
of the differential equation
\begin{equation}
\label{ode_chi}
  \left[ -\partial_{\bot}^2+k_{\|}^2+M_H(h)^2 \right]
  \psi(x_{\bot})=0,
\end{equation}  
with the boundary conditions 
\begin{equation}
\label{bc2}
  \psi(-L)=0; \quad  \psi(-L)'=1.
\end{equation}
The corresponding solution in the vacuum
sector, $\psi^{(0)}(x_{\bot})$,
can be taken from (\ref{psi_vac}):
\begin{equation}
\label{psi_vac2}
\psi^{(0)}(L)=\frac{ e^{(k_{\|}^2+ m^2)^{1/2} 2 L} }{
             2 (k_{\|}^2+ m^2)^{1/2} }.
\end{equation}

Note that Eq.~(\ref{sigma_chi}) must be rendered UV-finite
by the separation and subsequent dimensional regularization
of the UV-divergent piece.
At this point we only note that the regularization proceeds
according to the procedure established in section 2.
We will return to the actual treatment of ultraviolet
divergences in $\sigma_{\chi}$ later.

To solve the differential equation (\ref{ode_chi})
we divide the whole interval into three parts.
Namely, region I: $x_{\bot}<-a$, region II: $-a \le x_{\bot} \le a$,
and region III: $x_{\bot} \ge a$.
We can immediately write the general solution
of (\ref{ode_chi}) in regions I and III,
since $h(x_{\bot})$ restricted to these intervals
is constant.
Thus, we only need to integrate Eq. (\ref{ode_chi})
in region II,  and then match the solution at
the intersections $x_{\bot}=-a$ and $x_{\bot}=a$.

Omitting terms that fall off exponentially
with increasing $L$, the solution in region I
is given by
\begin{equation}
  \psi(x_{\bot})=\frac{1}{2 (k_{\|}^2+m^2)^{1/2} } \;
        e^{(k_{\|}^2+m^2)^{1/2} (L+x_{\bot})}, \qquad
          x_{\bot} \in [-L,a]
\end{equation} 
The solution in region II can be
written in the following form:
\begin{equation}
\label{psi_i2}
   \psi(x_{\bot})=\frac{
     f^{(II)}(x_{\bot}) }{2 (k_{\|}^2+m^2)^{1/2} } \;
        e^{(k_{\|}^2+m^2)^{1/2} (L+x_{\bot})},
      \qquad     x_{\bot} \in [-a,a]
\end{equation}
where $ f^{(II)}(x_{\bot}) $ satisfies
the equation
\begin{equation}
\label{f_2}
 \left[-\partial_{\bot}^2-2 (k_{\|}^2+ m^2 )^{1/2}
  \partial_{\bot} + \frac{\lambda}{2} h(x_{\bot})^2 
 \right]  f^{(II)}(x_{\bot})=0, 
    \qquad  x_{\bot} \in [-a,a]
\end{equation}
with the boundary conditions
\begin{equation}
 f^{(II)}(-a)=1; \quad { f^{(II)}(-a) }'=0.
\end{equation}
The general solution in region III is 
\begin{equation}
\label{psi3}
  \psi(x_{\bot})=f^{(III)} e^{(k_{\|}^2+m^2)^{1/2} L+
    (k_{\|}^2+ M_H(\upsilon)^2)^{1/2} x_{\bot} }+
    g^{(III)} e^{(k_{\|}^2+m^2)^{1/2} L-
    (k_{\|}^2+ M_H(\upsilon)^2)^{1/2} x_{\bot} }, \, \, \, x \in [a,L]
\end{equation}
Matching this solution with (\ref{psi_i2}) by
requiring the continuity of $\psi(x_{\bot})$
and its first derivative, we obtain the
following expression for  $f^{(III)}$
\begin{equation}
\label{f}
 f^{(III)}  =   \frac{ e^{a [(k_{\|}^2+ m^2 )^{1/2}-
   (k_{\|}^2+ M_H(\upsilon)^2)^{1/2}] } }{
  2  (k_{\|}^2+ m^2 )^{1/2} }    
\left( \frac{ { f^{(II)}(a) }' }{
   (k_{\|}^2{+}M_H(\upsilon)^2)^{1/2} }
 + f^{(II)}(a) \left[ 1 +
     \frac{(k_{\|}^2+m^2 )^{1/2}}{(k_{\|}^2{+}M_H(\upsilon)^2)^{1/2} }
  \right] \right).
\end{equation}

It is $f^{(III)}$ that determines the
value of $\sigma_{\chi}$, as the second
term in (\ref{psi3}) is exponentially
suppressed in $L$.
One can therefore take the first term in
(\ref{psi3}) with  $f^{(III)}$ given by (\ref{f})
and substitute this, together with $\psi^{(0)}(L)$
from (\ref{psi_vac2}) into the expression (\ref{sigma_chi})
for $\sigma_{\chi}$.
Before we do that, however, note that for
the constant background field 
$h(x_{\bot}) \equiv \upsilon$, $\psi(L)$ can
be easily found analytically.
This allows one to compute the part of the effective
action $\Gamma_{\chi}[h \equiv \upsilon]$ at finite  $L$:
\begin{equation}
\label{gamma_chi_upsilon}
\frac{ \Gamma_{\chi}[\upsilon] }{ V_{\|} }=\frac{1}{2} 
    \int \frac{ d^{n-1}k_{\|} }{ (2 \pi)^{n-1} } \left[
  2 L \left[ (k_{\|}^2+ M_H(\upsilon)^2)^{1/2}
    -(k_{\|}^2+m^2)^{1/2} \right]
  + \frac{1}{2} \ln \left( \frac{ k_{\|}^2+M_H(\upsilon)^2 }{
                  k_{\|}^2+m^2 } \right)
   \right].
\end{equation}
The term that multiplies $L$ in this expression comes
from the exponent in (\ref{psi_vac2}) and the corresponding
exponent in $\psi(L)$, while the ratio inside the $\log$ in
the last term involves the coefficients that multiply
the respective exponents.
One can obtain part of the effective potential
[the $M_H(\Phi)^3$ term in Eq.~(\ref{effective_potential})]
by dividing the expression above by the
volume of the spacetime box $2 V_{\|} L$ and taking
the limit $L \rightarrow \infty$. 
The last term in (\ref{gamma_chi_upsilon}) is a boundary
effect reflecting our choice of Dirichlet boundary
conditions.
It makes no contribution to the effective potential
in the $L \rightarrow \infty$ limit.

We may require the parameters
of the theory formulated in a box with finite length
$L$ to be adjusted so
that the effective action evaluated at finite $L$
vanishes at $\upsilon$.
Of course, as we send $L$ to infinity, these parameters
will flow to the corresponding values for the theory
in infinite spacetime.
This condition will prove useful momentarily.
For one can notice that the part of $\sigma_{\chi} V_{\|}$
that is linear in $L$ is simply half of the 
first term in (\ref{gamma_chi_upsilon}).
Thus, one can separate $\sigma_{\chi}$ as
\begin{equation}
\label{sigma_chi3}
\sigma_{\chi}=\tilde{\sigma}_{\chi}+\frac{
       \Gamma_{\chi}[\upsilon]}{2 V_{\|}},
\end{equation}
where
\begin{equation}
\label{sigma_chi_tilde}
\tilde{\sigma}_{\chi} =  \frac{1}{2} 
    \int \frac{ d^{n-1}k_{\|} }{ (2 \pi)^{n-1} } \left[
   \ln \left( \frac{ 2 f^{(III)} }{
                  (k_{\|}^2+m^2)^{1/2} } \right)
   -\frac{1}{4}
    \ln \left( \frac{ k_{\|}^2+M_H(\upsilon)^2 }{
                  k_{\|}^2+m^2 } \right)   
                  \right]
\end{equation}
is finite as $L \rightarrow \infty$ and 
$\Gamma_{\chi}[\upsilon]$ is given by Eq. (\ref{gamma_chi_upsilon}).

Recall that 
$\Gamma_{\chi}[\upsilon]$  is a part of the effective
action evaluated at the constant field configuration 
$h(x_{\bot})=\upsilon$.
One can anticipate that the same story will
repeat for the calculation of $\sigma_{A\phi}$,
and $\Gamma_{A \phi}[\upsilon]$ will be
produced in a similar fashion.
Thus, after we collect all parts of $\sigma$
together, the coefficient in front of $L$ in the
resulting expression will be a multiple of
$\Gamma[\upsilon]$, and therefore the divergence
in $L$ will vanish, leaving an expression for
$\sigma$ that is finite in the limit $L \rightarrow \infty$.

Having anticipated the form of the outcome, let us now compute
$\sigma_{A \phi}$.
The relevant differential operators 
in (\ref{gamma_aphi}) are now given by
$2 \times 2$ matrices, so we must find a way to
generalize the correspondence between determinants
and solutions of associated differential equations.
This is done in the appendix, where we show that
\begin{equation}
\label{det_ntwo}
\frac{
 \det \, \tilde{\Delta}_{A \phi}}{
 \det \, \tilde{\Delta}_{A \phi}^{(0)} }=
  \frac{
   \psi_1^{(1)}(L) \psi_2^{(2)}(L)-
     \psi_2^{(1)}(L) \psi_1^{(2)}(L) }
    {  \varphi_1^{(1)}(L) \varphi_2^{(2)}(L)-
     \varphi_2^{(1)}(L) \varphi_1^{(2)}(L) },
\end{equation}   
where $\psi^{(1)}(x_{\bot})$ and 
$\psi^{(2)}(x_{\bot})$ are solutions 
of the same system of differential
equations,
\begin{eqnarray}
   \tilde{\Delta}_{A \phi} \;
 \left(  \begin{array}{c}
   \psi_1^{(i)}(x_{\bot}) \\
   \psi_2^{(i)}(x_{\bot})
   \end{array} \right) \;=0, \qquad i=1, \;2,
\end{eqnarray}
with vanishing initial conditions,
\begin{eqnarray}
   \left(  \begin{array}{c}
   \psi_1^{(i)}(-L) \\
   \psi_2^{(i)}(-L)
   \end{array} \right) \;=
   \left(  \begin{array}{c}
   0 \\
   0
   \end{array} \right), \qquad i=1, \; 2,
\end{eqnarray}
but with differing initial derivatives,
\begin{eqnarray}
  \left(  \begin{array}{c}
   {\psi_1^{(1)}}'(-L) \\
   {\psi_2^{(1)}}'(-L)
   \end{array} \right) \;=
   \left(  \begin{array}{c}
   1 \\
   0
   \end{array} \right),
\qquad
 \left(  \begin{array}{c}
   {\psi_1^{(2)}}'(-L) \\
   {\psi_2^{(2)}}'(-L)
   \end{array} \right) \;=
   \left(  \begin{array}{c}
   0 \\
   1
   \end{array} \right).
\end{eqnarray}
In other words, $\psi^{(1)}$ and  $\psi^{(2)}$ are
the two linearly independent solutions of 
$\tilde{\Delta}_{A \phi} \psi^{(i)}=0$ which vanish at
$x_{\bot}=-L$ but which have orthogonal initial
``velocities''.

The corresponding vacuum sector solutions
are denoted by  $\varphi^{(i)}$.

Hence, $\sigma_{A\phi}$ can be written 
analogously to (\ref{sigma_chi}) as
\begin{equation}
\sigma_{A\phi}=\frac{1}{2} \int 
       \frac{d k^{n-1}_{\|}}{ (2 \pi)^{n-1} }
       \ln \left( \frac{
   \psi_1^{(1)}(L) \psi_2^{(2)}(L)-
     \psi_2^{(1)}(L) \psi_1^{(2)}(L) }
    {  \varphi_1^{(1)}(L) \varphi_2^{(2)}(L)-
     \varphi_2^{(1)}(L) \varphi_1^{(2)}(L) } \right).
\end{equation}

The further computation of $\sigma_{A \phi}$ 
proceeds similarly to that of $\sigma_{\chi}$
in almost all respects.
We again divide the interval
$[-L,L]$ into three regions, and use appropriate changes 
of variables to extract the asymptotic
behavior of the solutions.
The main results are presented below.

As in the case of $\sigma_{\chi}$, $\sigma_{A\phi}$
can be decomposed analogously to (\ref{sigma_chi3}):
\begin{equation}
\label{sigma_aphi3}
\sigma_{A \phi}=\tilde{\sigma}_{A \phi}+\frac{
       \Gamma_{A \phi}[\upsilon]}{2 V_{\|}},
\end{equation}
where
\begin{eqnarray}
\label{sigma_aphi_tilde}
\nonumber
 \sigma_{A \phi}    &=& \displaystyle
   \frac{1}{2}  \int \frac{ d^{n-1}k_{\|} }
  { (2 \pi)^{n-1} } \Bigg[ \ln  \left( 4 \frac{
      f_1^{(1,III)}  f_2^{(2,III)}{-}
                f_1^{(2,III)}  f_2^{(1,III)} }{
      (k_{\|}^2+m^2)^{1/2} \, k_{\|}  }  \right)
 \\ && \displaystyle \qquad \qquad
   {-}\frac{1}{4}
    \ln  \left( \frac{ (k_{\|}^2{+}M_W(\upsilon)^2)\,
                 (k_{\|}^2{+}M_{A\phi}(\upsilon)^2) }{
                k_{\|}^2 \, (k_{\|}^2+m^2) } \right)
 \Bigg] 
\end{eqnarray}
and
\begin{eqnarray}
\label{gamma_aphi_upsilon}
\nonumber \displaystyle
\frac{\Gamma_{A \phi}[\upsilon]}{V_{\|}} &=& \frac{1}{2} 
    \int \frac{ d^{n-1}k_{\|} }{ (2 \pi)^{n-1} } \Bigg[
  2 L \left\{ (k_{\|}^2+M_W(\upsilon)^2)^{1/2}-k_{\|} +
    (k_{\|}^2+ M_{A\phi}(\upsilon)^2)^{1/2}-
    (k_{\|}^2+ m^2)^{1/2}  \right\} \\ 
   && \displaystyle \qquad \qquad \qquad
 +\frac{1}{2} \ln \left( \frac{ (k_{\|}^2+M_W(\upsilon)^2)
                 \, (k_{\|}^2+M_{A\phi}(\upsilon)^2) }{
                k_{\|}^2 \, (k_{\|}^2+m^2) } \right)
  \Bigg],
\end{eqnarray}
where $f_i^{(j,III)}$ is determined
analogously to $f^{(III)}$ in (\ref{f}). 
The subscript $i$ in the
former labels the component of a vector,
while the superscript $j$ distinguishes the
two sets of boundary conditions.
In other words, $f_i^{(j,III)}$ comes from
the $\psi_i^{(j)}(x_{\bot})$.

Explicit expressions for $f_i^{(1,III)}$ are:
\begin{eqnarray}
\label{f_i1}
\nonumber f_1^{(1,III)}   &=&    
\frac{ e^{a [ k_{\|}-
  (k_{\|}^2+M_W(\upsilon)^2)^{1/2} ] }  }{
    2  k_{\|}  } \left[ 
 \frac{  {f_1^{(1,II)}}'(a) }{
   (k_{\|}^2+ M_W(\upsilon)^2)^{1/2} }
  {+} f_1^{(1,II)}(a) \left( 1 {-}
     \frac{ k_{\|} }{
  (k_{\|}^2+ M_W(\upsilon)^2 )^{1/2} }
 \right) \right], \\ \\ \nonumber
f_2^{(1,III)}  & =&    
\frac{ e^{a [ k_{\|}-
  (k_{\|}^2+M_{A\phi}(\upsilon)^2)^{1/2} ] }  }{
    2  k_{\|}  } \left[
 \frac{ { f_2^{(1,II)}}'(a) }{
   (k_{\|}^2+ M_{A\phi}(\upsilon)^2)^{1/2} }
  {+} f_2^{(1,II)}(a) \left( 1 {-}
     \frac{ k_{\|} }{
  (k_{\|}^2+ M_{A\phi}(\upsilon)^2)^{1/2} }
 \right) \right], \\
\end{eqnarray}
where $f_1^{(1,II)}(x_{\bot})$ and 
$f_2^{(1,II)}(x_{\bot})$ are the solutions
of the following system of ODE's,
\begin{eqnarray}
\label{ode_complicated1}
\left( \begin{array}{ccc}
   -\partial_{\bot}^2 -2 k_{\|} \partial_{\bot}
       + M_W(h)^2
   & -2 g (\partial_{\bot} h)\\
   -2 g (\partial_{\bot} h)
 &  -\partial_{\bot}^2 -2 k_{\|} \partial_{\bot}+
   M_{A\phi}(h)^2
\end{array} \right)
  \left(  \begin{array}{c}
   f_1^{(1,II)} \\
   f_2^{(1,II)})
   \end{array} \right) \;=0,
\end{eqnarray}
subject to the boundary conditions
\begin{eqnarray}
  \left(  \begin{array}{c}
   f_1^{(1,II)}(-a) \\
   f_2^{(1,II)}(-a)
   \end{array} \right) \;=
   \left(  \begin{array}{c}
   1 \\
   0
   \end{array} \right), \qquad
  \left(  \begin{array}{c}
   {f_1^{(1,II)}}'(-a) \\
   {f_2^{(1,II)}}'(-a)
   \end{array} \right) \;=
   \left(  \begin{array}{c}
   0 \\
   0
   \end{array} \right).
\end{eqnarray}

Corresponding expressions for $f_i^{(2,III)}$ are
very similar, as one might expect.
The only difference is caused by the fact that
$\psi^{(1)}(x_{\bot})$ and $\psi^{(2)}(x_{\bot})$
have different exponents in the region I.
One finds,
\begin{eqnarray}
\label{f_i2}
\nonumber
f_1^{(2,III)}  & = &   
\frac{ e^{a [ (k_{\|}^2+m^2)^{1/2}-
  (k_{\|}^2+M_W(\upsilon)^2)^{1/2} ] }  }{
    2  (k_{\|}^2+m^2)^{1/2}  } \left[ 
 \frac{  {f_1^{(2,II)}}'(a) }{
   (k_{\|}^2{+}M_W(\upsilon)^2)^{1/2} }
  {+}f_1^{(2,II)}(a) \left( 1 {-}
     \frac{  (k_{\|}^2+m^2)^{1/2} }{
  (k_{\|}^2{+}M_W(\upsilon)^2 )^{1/2} }
 \right) \right], \\ \\ \nonumber
f_2^{(2,III)}  & = &   
\frac{ e^{a [  (k_{\|}^2+m^2)^{1/2}-
  (k_{\|}^2+M_{A\phi}(\upsilon)^2)^{1/2} ] }  }{
    2  k_{\|}  } \left[
 \frac{ { f_2^{(2,II)}}'(a) }{
   (k_{\|}^2{+}M_{A\phi}(\upsilon)^2)^{1/2} }
 {+} f_2^{(2,II)}(a) \left( 1 {-}
     \frac{  (k_{\|}^2+m^2)^{1/2} }{
  (k_{\|}^2{+}M_{A\phi}(\upsilon)^2)^{1/2} }
 \right) \right], \\
\end{eqnarray}
where $f_1^{(2,II)}(x_{\bot})$ and 
$f_2^{(2,II)}(x_{\bot})$ are the solutions
of the following system of ODE's,
\begin{eqnarray}
\label{ode_complicated2}
\left( \begin{array}{ccc}
  -\partial_{\bot}^2 {-}2 (k_{\|}^2{+}m^2)^{1/2}
       \partial_{\bot}{-}m^2
       {+} M_H(h)^2 
   &-2 g (\partial_{\bot} h)\\
   -2 g (\partial_{\bot} h) 
 &  -\partial_{\bot}^2 {-}2 (k_{\|}^2{+}m^2)^{1/2}
   \partial_{\bot}{-}m^2 {+} M_{A\phi}(h)^2
\end{array} \right)
  \left(  \begin{array}{c}
   f_1^{(2,II)} \\
   f_2^{(2,II)}
   \end{array} \right)=0, \kern-45pt
\nonumber\\
\end{eqnarray}
subject to the boundary conditions
\begin{eqnarray}
  \left(  \begin{array}{c}
   f_1^{(2,II)}(-a) \\
   f_2^{(2,II)}(-a)
   \end{array} \right) \;=
   \left(  \begin{array}{c}
   0 \\
   1
   \end{array} \right), \qquad
  \left(  \begin{array}{c}
   {f_1^{(2,II)}}'(-a) \\
   {f_2^{(2,II)}}'(-a)
   \end{array} \right) \;=
   \left(  \begin{array}{c}
   0 \\
   0
   \end{array} \right).
\end{eqnarray}

Now one can collect together $\sigma_{\rm cl}$,
$\sigma_{\chi}$ and $\sigma_{A\phi}$, and in particular
combine the terms which add up to make
$\Gamma[\upsilon]/2 V_{\|}$, and set
it to zero.
At this point we need to recall our prescription
for regularization, which was introduced in section 2.
Namely, we need to subtract the divergent pieces from 
$\sigma_{\chi}$ and $\sigma_{A\phi}$, and then add 
them back, with the divergent loop diagrams replaced by
their regularized values.

Naively, the divergent term in $\sigma_{\chi}$,
for example, is given by
\begin{equation}
\label{uv_divergence}
\sigma_{\chi}^{\rm (div)}= \frac{\lambda}{4} G \; {\rm tr} \left[
          (\nabla^2+m^2)^{-1} \right],
\end{equation}
where
\begin{equation}
  G =\int_{-L}^L dx_{\bot} h^2(x_{\bot})
\end{equation}
is defined analogously to $F$ in Eq.~(\ref{P}),
and the trace involves integration over
$k_{\|}$ and summing over discrete $x_{\bot}$ modes
that vanish at $\pm L$.
However, when extracting 
$\Gamma[\upsilon]/2 V_{\|}$,
we should also add and subtract the corresponding
UV-divergent term which is just
\begin{equation}
\frac{\lambda}{4} L \upsilon^2 \; {\rm tr} 
             \left[ (\nabla^2+m^2)^{-1} \right]=
  \frac{\lambda}{4} \left( \int_0^L dx_{\bot} \upsilon^2 
  \right)\;
      {\rm tr} 
             \left[ (\nabla^2+m^2)^{-1} \right].
\end{equation}
Therefore, the UV-subtraction in $\tilde{\sigma}_{\chi}$
should actually involve $\sigma_{\chi}^{\rm (div)}$ as in
Eq. (\ref{uv_divergence}) but with $G$
replaced by
\begin{equation}
\label{uv_divergence2}
\tilde{G}= \int_{-L}^L dx_{\bot} \left[  h^2(x_{\bot})-
                                \theta(x_{\bot}) \upsilon^2 \right].
\end{equation}
Here $\theta(x_{\bot})$ is the step function
which is zero for $x_{\bot}<0$ and one for  $x_{\bot}>0$.
Note that $\tilde{G}$ is finite in the
limit $L \rightarrow \infty$.

The treatment of ultraviolet divergences in
$\tilde{\sigma}_{A \phi}$ proceeds in a similar
manner.

Taking the $L \rightarrow \infty$ limit we arrive at
an expression for $\sigma$ that only involves
finite integrals that can be computed numerically:

\begin{equation}
\label{sigma_main_result}
 \sigma=  \lim_{L \rightarrow \infty} \left[ \tilde{\sigma}_{\rm cl}+
     \tilde{\sigma}_{\chi}+\tilde{\sigma}_{A \phi}
  \right],
\end{equation}
where 
\begin{eqnarray}
\label{sigma_cl_tilde}     
\tilde{\sigma}_{\rm cl} &=& \displaystyle \int dx_{\bot} \left\{
  \frac{1}{2}  (\partial_{\bot} h(x_{\bot}) )^2+
      \frac{m^2}{2}  \left[ h(x_{\bot})^2-
                        \theta(x_{\bot}) \upsilon^2 \right]
        +\frac{\lambda}{4!} \left[ h(x_{\bot})^4-
                    \theta(x_{\bot}) \upsilon^4 \right]
          \right\}, \\ 
\label{sigma_chi_tilde_final}
\nonumber
\tilde{\sigma}_{\chi} &=&\displaystyle  \frac{1}{2} 
    \int \frac{ d^2k_{\|} }{ (2 \pi)^2 } \left[
   \ln \left( \frac{ 2 f^{(III)} }{
                  (k_{\|}^2{+}m^2)^{1/2} } \right)
   -\frac{1}{4}
    \ln \left( \frac{ k_{\|}^2{+}M_H(\upsilon)^2 }{
                  k_{\|}^2+m^2 } \right)
   -\frac{ \lambda}{4}\frac{\tilde{G}}{(k_{\|}^2{+}m^2)^{1/2}}
                  \right]
  \\  && \displaystyle
  +\frac{ \lambda}{4}\frac{\tilde{G}}{4 \pi}m, \\ 
\label{sigma_aphi_tilde_final}
\nonumber
 \tilde{\sigma}_{A \chi} &=& \displaystyle
   \frac{1}{2}  \int \frac{ d^2k_{\|} }
  { (2 \pi)^2 } \Bigg[ \ln \left( 4 \frac{
      f_1^{(1,III)}  f_2^{(2,III)}{-}
                f_1^{(2,III)}  f_2^{(1,III)} }{
      (k_{\|}^2+m^2)^{1/2} k_{\|}  } \right)
  \\ \nonumber && \displaystyle
  \qquad \qquad -\frac{1}{4}
    \ln  \left( \frac{ (k_{\|}^2{+}M_W(\upsilon)^2)
                 (k_{\|}^2{+}M_W(\upsilon^2)) }{
                k_{\|}^2(k_{\|}^2+m^2) } \right)
  -\frac{g^2}{2} \frac{\tilde{G}}{k_{\|}} \;
 - (\frac{ \lambda}{12}+\frac{g^2}{2}) 
      \frac{\tilde{G}}{(k_{\|}^2+m^2)^{1/2}}
 \Bigg]  
  \\  && \displaystyle
  +(\frac{ \lambda}{12}+\frac{g^2}{2}) 
      \frac{\tilde{G}}{4 \pi}m. 
\end{eqnarray}

Note that the ``classical'' term (\ref{sigma_cl_tilde})
is also finite as $L \rightarrow \infty$, as it is
defined by the same subtraction that
was needed to isolate $\Gamma_{\rm cl}[\upsilon]$,
\begin{equation}
 \tilde{\sigma}_{\rm cl}=\sigma_{\rm cl}-\frac{\Gamma_{\rm cl}[\upsilon]}{
                      2 V_{\|} }.
\end{equation}
The terms outside the integrals in $\tilde{\sigma}_{\chi}$
and $\tilde{\sigma}_{A\phi}$ are the values
of UV-divergent terms of the type (\ref{uv_divergence}),
computed in the limit of infinite $L$ by dimensional
continuation.
%
%
%
%
%
%
%
%
%
%
%
\subsection{Numerical evaluation}

It is now possible, by using 
Eq's.~(\ref{sigma_main_result})--(\ref{sigma_aphi_tilde_final}),
to compute the full one-loop surface tension
for an arbitrary domain wall-type field configuration.
One just needs to solve the relevant differential
equations (\ref{f_2}), (\ref{ode_complicated1}),
and (\ref{ode_complicated2}), and then perform a
final integration
over the worldvolume momentum $k_{\|}$.
The latter step merely involves a one-dimensional integration
over the radial component of $k_{\|}$, as the
integrands in $\sigma_{\chi}$ and $\sigma_{A\phi}$
depend only on $k_{\|}^2$.

We have written a simple {\it Mathematica} program 
to implement these steps.
It allowed computation of $\sigma$ with a relative 
accuracy of $0.2\%$ or better.\footnote{
Higher accuracy is undoubtedly feasible given 
more care than we used in the numerical implementation.
This is not completely trivial however, as
Eq's. (\ref{ode_complicated1}) and 
(\ref{ode_complicated2}) can be quite stiff due to the
presence of two different length
scales set by the mass parameters $M_W(h)$ and 
$M_{A\phi}(h)$.}
As described below, we tested the numerical results against 
perturbative calculations when $\lambda/g^2<<1$,
and obtained good agreement.

In principle, one may calculate the one-loop energy
of the true domain wall configuration by minimizing 
(\ref{sigma_main_result})
with respect to the domain wall profile.
This we have not bothered to do.

%
%
%
%
%
%
%
%
%
%

\subsection{Comparison with perturbative $\lambda/g^2$
            expansion}

As noted earlier, for our specific example of the Abelian Higgs
model, the ratio of scalar to gauge couplings,
$\lambda/g^2$, functions as the loop expansion parameter
controlling the reliability of perturbation theory
at the phase transition.
For asymptotically small values of $\lambda/g^2$ one
may analytically compute the leading and first subleading
correction to the surface tension $\sigma$.
One way to do this is to start with the derivative
expansion of the effective action:
\begin{equation}
\label{Gamma_dexpansion}
\Gamma = \int d^3x \left[ 
  V_{\rm eff}(h)+\frac{1}{2} Z(h) (\partial_\mu h)^2+
      \cdots \right],
\end{equation}
where $\cdots$ stands for the terms with four or more 
derivatives.
We denote the first two terms in this expansion
by $\Gamma^{(2)}$.

To leading order in $\eta \equiv \lambda/g^2$
one may take the wavefunction renormalization factor $Z(h)$
in the effective action (\ref{Gamma_dexpansion})
to equal one.
The reason for this is that the one-loop
corrections to the classical 
kinetic term in (\ref{Gamma_dexpansion})
behave like $g^4 h^2 M_W(h)^{-3} (\partial h)^2$
and $\lambda^2 h^2 M_H(h)^{-3} (\partial h)^2$,
and hence their contribution to $\sigma$ is suppressed in 
comparison with the leading term by a factor
of $g^2/M_W(\upsilon) \sim \eta$ and 
$\lambda/M_H(\upsilon) \sim \eta^{3/2}$,
respectively.
(Recall, from (\ref{relations_atv}), that 
$M_W(\upsilon)=g \upsilon \sim g^4/\lambda$
while $M_H(\upsilon) \sim m \sim g^3/\sqrt{\lambda}$.)
Hence, the leading term in the expansion of
$\sigma$ is completely determined by the 
leading term in the expansion of the one-loop 
effective potential in powers of $\eta$.
Recall that for a scalar theory with
a canonical kinetic term and any effective
potential with two degenerate minima at $\Phi=0$
and $\Phi=\upsilon$, normalized so that 
$ V_{\rm eff}(0)= V_{\rm eff}(\upsilon)=0$,
the domain wall field configuration is a
solution of the classical equations of motion
whose conserved ``energy'' 
$E \equiv \frac{1}{2}(\partial_{\bot}h)^2-V_{\rm eff}(h)$
vanishes.
Consequently, the surface tension, neglecting the
wavefunction renormalization and all higher derivative
terms, is given by
\begin{equation}
\label{sigma_int}
 \sigma = \int_{- \infty}^{\infty} dx_{\bot} \; (\partial_{\bot} h)^2= 
  \int_0^{\upsilon} d h \; \sqrt{ 2 V_{\rm eff}(h)}.
\end{equation}

To obtain the leading behavior of the integral
(\ref{sigma_int}), one may expand the one-loop
potential (\ref{effective_potential}) for small
$\eta$, obtaining just the classic double well form
(\ref{veff_double_well}).
Inserting this into Eq. (3.70) gives
\begin{equation}
\sigma= \sigma^{(0)} \left(1+ {\cal O}(\eta) \right),
\end{equation}
with
\begin{equation}
\label{sigma_zero}
 \sigma^{(0)} = \frac{m}{\upsilon} \int_0^{\upsilon} dh \;
           h(\upsilon-h)= \frac{1}{6} m \upsilon^2=
           \frac{2}{\sqrt{3} \pi} m^2 \eta^{-3/2}.
\end{equation}
[The final form uses the leading behavior of
$\upsilon^2=\frac{4 \sqrt{3}}{\pi}\;  \eta^{-3/2} \; m \; (1+{\cal O}(\eta))$.]
To get the complete next-to-leading term, it is necessary
to calculate both the two-loop effective potential and 
the one-loop correction to the wavefunction renormalization,
{\it i.e.}, $Z(h)-1$.
One can not go beyond  
the next-to-leading order term using the derivative
expansion, since the Higgs field contributions to
the higher order terms in the expansion (\ref{Gamma_dexpansion})
give corrections to the surface tension that
all scale like $\sigma^{(0)} \eta^{3/2}$.
To see this, one may look, for example, at the contributions
to the two- and four-derivative terms that 
behave like
$\lambda^2 h^2 M_H(h)^{-3} (\partial h)^2$ and
$\lambda^2 M_H(h)^{-5} (\partial h)^4$, respectively.
The first term represents a $Z$-factor correction and
scales like $\sigma^{(0)} \eta^{3/2}$.
The second term is further suppressed by a factor
of $[l M_H(\upsilon)]^{-2}$ where $l$
is the characteristic thickness of domain wall.
But this is ${\cal O}(M_H(\upsilon)^{-1})$ and so
this four derivative contribution also scales
like $\sigma^{(0)} \eta^{3/2}$.
It is easy to see that further higher order derivative
terms also scale like  $\sigma^{(0)} \eta^{3/2}$.

More generally, the expansion of $\sigma$ in powers of $\eta$
breaks down because of infrared
problems \cite{Laser}.
This can be seen by looking at the contribution
of the gauge field to the four derivative term
computed to one-loop order.
This term behaves like $g^4 M_W(h)^{-5} (\partial h)^4$
and by naive power counting
gives a contribution to $\sigma$ that
scales as $\sigma^{(0)} \eta^2$.
However, since $h(x_{\bot})$ vanishes exponentially in
the symmetric phase ({\it i.e.}, as $x_{\bot} \rightarrow -\infty$),
and $M_W(h)=g h$, the integral over $x_{\bot}$ in the effective
action diverges.

In fact, there is another, more significant source
of infrared divergences.
One can see this by expanding the effective potential to
higher orders in $\eta$.
One finds that the integral which enters
the ${\cal O}(\eta^{3/2})$ term is logarithmically divergent 
in the infrared.
In particular, this implies that the
next-to-next-to-leading order correction to 
$\sigma$ is ${\cal O}(\eta^{3/2}\log \eta)$.

Nevertheless for small $\lambda/g^2$, one might hope that the one-loop
value of $\sigma$ could be well approximated
by the contributions from the effective potential 
plus the first wavefunction renormalization correction,
while neglecting all higher order terms in the
derivative expansion.
It was argued in Ref. \cite{Laser} that this is 
indeed the case in electroweak theory, at least for
intermediate values of the Higgs masses.
Unfortunately the infrared divergences of the
individual terms in the various
expansions used in  \cite{Laser} complicate
the accurate computation of $\sigma$.
In contrast, the technique  that we use
in this paper does not have
infrared problems, as it effectively resums all
one-loop corrections to the surface tension,
allowing its accurate computation for any value of
$\lambda/g^2$.

In what follows we show
that the inclusion of the wavefunction renormalization
gives results that are very close to the full one-loop value 
of $\sigma$ given by (\ref{sigma_main_result}),
as long as $\lambda/g^2$ is less than $1$.
The comparison with analytic results  also 
provides a nice consistency check
of Eq. (\ref{sigma_main_result}). 

To make this comparison, it is necessary to
calculate the wavefunction renormalization 
in the particular gauge (\ref{gauge_fix}).
We use the convenient method of Ref. \cite{Caro&Salcedo}
to determine the first two terms in the derivative
expansion of the effective action
(\ref{Gamma_dexpansion}), which we denote by
$\Gamma^{(2)}$.
According to Ref. \cite{Caro&Salcedo}, the one-loop 
contribution to $\Gamma^{(2)}$ can be written as
\begin{equation}
\delta \Gamma^{(2)}_{\rm one-loop}=
\delta \Gamma^{(2)}_{\chi}+\delta \Gamma^{(2)}_{A\phi},
\end{equation}
where
\begin{equation}
\delta \Gamma^{(2)}_{\chi}= \frac{1}{2}
 \int d^3 x \int \frac{d^3 k}{(2 \pi)^3}
 \left[  \ln \frac{ {\hat \Delta}_{\chi}}{
                {\hat \Delta}_{\chi}^{(0)} }
   +\frac{k^2}{3} \left( 
      \partial_{\mu} {\hat \Delta}_{\chi}^{-1}
                \right)^2 \right],
\end{equation}
and
\begin{equation}
\label{Gamma2_aphi}
\delta \Gamma^{(2)}_{A\phi}=
 \frac{1}{2} \int d^3 x \int \frac{d^3 k}{(2 \pi)^3}
 \left[
   {\rm tr} \; \ln \frac{ {\hat \Delta}_{A\phi}}{
                {\hat \Delta}_{A\phi}^{(0)} }
   +\frac{k^2}{3} {\rm tr} \left( 
      \partial_{\mu} {\hat \Delta}_{A\phi}^{-1}
                \right)^2
 \right].
\end{equation}
In the equations above, the trace runs over matrix
indices and the hat on top of
the operators $\Delta_{\chi}$ and $\Delta_{A\phi}$
[defined in Eq's. (\ref{Delta_chi}) and (\ref{Delta_aphi})]
denotes a Fourier transform with respect to $k$.
For example,
\begin{equation}
   {\hat \Delta}_{\chi}=k^2+M_H(h)^2.
\end{equation}
The wavefunction renormalization 
\begin{equation}
\label{z_factor}
Z(h)=1+\delta Z_{\chi}+ \delta Z_{A\phi}
\end{equation}
receives contributions both from
$\delta \Gamma^{(2)}_{\chi}$ and  
$\delta \Gamma^{(2)}_{A\phi}$, denoted by
$\delta Z_{\chi}$ and $\delta Z_{A\phi}$, respectively.
The scalar contribution can be easily evaluated:
\begin{equation}
\delta Z_{\chi}= \frac{\lambda^2 h^2}{3}
    \int \frac{d^3 k}{(2 \pi)^3}
  \frac{k^2}{(k^2+M_H(h)^2)^4}=\frac{\lambda^2 h^2}{M_H(h)^3}
    \frac{1}{192 \pi}.
\end{equation}
To calculate $\delta \Gamma^{(2)}_{A\phi}$
one should notice that contributions with
two derivatives come from both terms in Eq. (\ref{Gamma2_aphi}),
as the fluctuation operator (\ref{Delta_aphi})
already contains the derivative term 
$g (\partial h)$.
A computation which is analogous to the one done
in \cite{Laser} leads to the result,
\begin{equation}
\label{z_aphi}
  \delta Z_{A\phi}=-\frac{g^2}{M_W(h)+M_{A\phi}(h)}
                                         \frac{1}{\pi}
  +\left(\frac{g^4 h^2}{M_W(h)^3}+
	 \frac{(\frac{\lambda}{6}+g^2)^2 h^2}{M_{A\phi}(h)^3}
   \right) \frac{1}{48 \pi}.
\end{equation}
One can easily check that $\delta Z_{A\phi} \sim \eta$ and 
$\delta Z_{\chi} \sim \eta^{3/2}$, in accord with
our earlier discussion.

The expression for the surface tension which
takes into account the $Z$-factor is a simple
generalization of (\ref{sigma_int}).
The vanishing ``energy'' now implies that
$\frac{1}{2} Z(h) (\partial_{\bot} h)^2=V_{\rm eff}(h)$, and
expression (\ref{sigma_int}) becomes
\begin{equation}
\label{sigma_z}
 \sigma =   \int_{- \infty}^{\infty} dx_{\bot} \; Z(h) 
   (\partial_{\bot} h)^2=
  \int_0^{\upsilon} dh \; \sqrt{ 2 Z(h) 
    V_{\rm eff}(h)}.
\end{equation}
We can compute the first two terms in the
expansion of this $Z$-factor corrected $\sigma$
by plugging in expressions (\ref{z_factor})--(\ref{z_aphi})
for $Z(h)$, and (\ref{effective_potential}) for
$V_{\rm eff}(h)$, rescaling the field as $h=\upsilon {\tilde h}$
to extract a factor of $\upsilon$, and expanding the
integrand in powers of 
$\eta$.
The leading term in the expansion is just $\sigma^{(0)}$.
To get the subleading term it is necessary to use
values of $g$ and $\upsilon$ accurate to
next-to-leading order:
\begin{eqnarray}
 g&=&3^{1/4} \sqrt{\pi} \eta^{1/4} m^{1/2} \left(1-\frac{3}{32} \eta
                      +{\cal O}(\eta^{3/2}) \right), \\
 \upsilon &=& \left( \frac{48}{\pi^2} \right)^{1/4} 
   \eta^{-3/4} m^{1/2} \left(1-\frac{1}{32} \eta
                      +{\cal O}(\eta^{3/2}) \right).
\end{eqnarray}
One finds that the first subleading term is given by
\begin{equation}
\sigma^{(1)}= \frac{m^2}{8 \sqrt{3} \pi} \eta^{-1/2}
   \int_0^1 d{\tilde h} \; ({\tilde h}-1)(17+18 {\tilde h})=
    -\frac{23 \; m^2}{16 \sqrt{3} \pi} \eta^{-1/2}.
\end{equation}
Hence, the $Z$-factor
corrected surface tension (\ref{sigma_z}) has the
small $\eta$ expansion
\begin{equation}
\label{sigma_z_exp}
\sigma  = 
   \sigma^{(0)} \left[1- \frac{23}{32} \eta +
      {\cal O}(\eta^{3/2} \; \log \eta) \right],
\end{equation}  
with $\sigma^{(0)}$ given in Eq. (\ref{sigma_zero}).

The result (\ref{sigma_z_exp}) is not the complete
next-to-leading order surface tension, as two-loop
corrections to the effective potential also contribute to 
the subleading ${\cal O}(\eta)$ term.
Including these two-loop contributions is straightforward,
but will not be examined here.
\begin{figure}[t]
\begin{center}
\setlength {\unitlength} {1cm}
\begin{picture}(0,0)
  \put(-1,2.0){$1-\sigma/\sigma^{(0)}$}
  \put(4.75,-0.5){$\lambda/g^2$}
\end{picture}
\def\epsfsize #1#2{1.0#1}
\epsfbox[75 0 363 163]{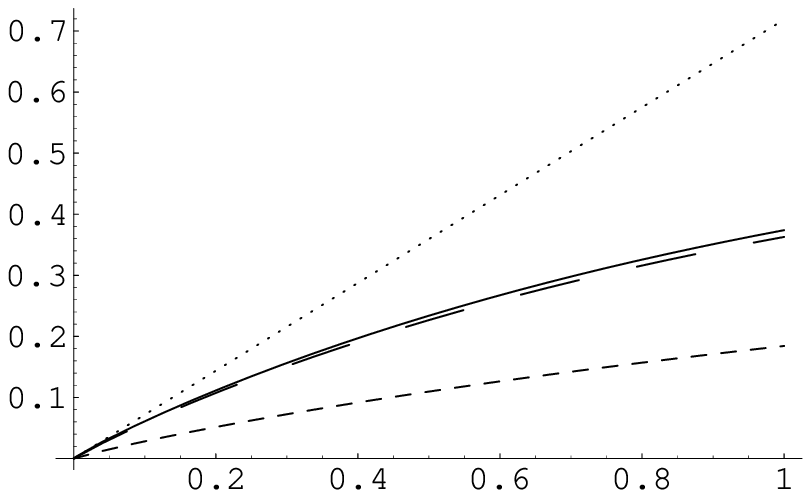}
\vspace{0.75cm}
\end{center}
\caption{ Comparison of the full one-loop surface tension
      (\ref{sigma_main_result}) [solid line], the second order
      derivative expansion result (\ref{sigma_z}) computed
      with the full one-loop effective potential and wavefunction
      renormalization [long dashes], the ``classical'' result
      (\ref{sigma_int}) computed with the full one-loop
      effective potential [short dashes], and the first two
      terms of the small $\eta$ asymptotic expansion (\ref{sigma_z_exp})
      [dotted line].
\label{fig1} }
\end{figure}

It is interesting to compare the different approximations
to the surface tension given by 
({\it i})
the full one-loop result (\ref{sigma_main_result}), 
({\it ii}) 
the expression (\ref{sigma_z}) computed with the
complete one-loop effective potential
 (\ref{effective_potential})
and wavefunction renormalization factor (\ref{z_factor}), 
({\it iii}) 
the expression (\ref{sigma_int}) computed with the
complete  one-loop effective potential (\ref{effective_potential}), 
and ({\it iv})
the first two terms of the
small $\eta$ asymptotic expansion (\ref{sigma_z_exp}). 
All these expressions omit two (and higher) loop contributions
to the effective potential, while ({\it ii}) also
omits four derivative (and higher)
contributions to the one-loop effective action, 
and ({\it iii}) omits all one-loop derivative 
corrections to the effective action.
This comparison is shown in Fig. \ref{fig1}.
The full one-loop result (\ref{sigma_main_result}) was
evaluated using the lowest order profile for the domain wall,
\begin{equation}
\label{field_profile}
 h(x_{\bot})=\frac{\upsilon}{2} \left[
        \tanh(m x_{\bot}/2)+1 \right],
\end{equation}
with $\upsilon$ being the true (not just leading order)
scalar vacuum expectation value in the Higgs phase.
Corrections to the profile of the domain wall only 
affect $\sigma$ at next-to-next to leading order
in the small $\eta$ expansion (\ref{sigma_z_exp}).
(This is tantamount to using the unperturbed wavefunction
to compute the first order correction to the 
energy of a state in quantum mechanics.)

>From Fig.~\ref{fig1} it is clear that the
difference between the second order derivative expansion result 
(\ref{sigma_z}) and the full one-loop result is quite
small, at least for $\lambda/g^2 \le 1$.
Numerically, this difference appears to scale as
$\eta^{3/2}$.
In contrast, the ``classical'' result (\ref{sigma_int})
which omits all one-loop derivative corrections, and the
leading two terms of the asymptotic expansion  (\ref{sigma_z_exp}),
both depart significantly from the full one-loop
result for quite small values of $\eta$.

%
%
%
%
%
%
%
%
%
%
%
%
%
%
%
%
\section{Conclusion}
The method presented in this paper provides a straightforward
technique for evaluating the one-loop energy density of any
configuration which depends non-trivially on only a
single coordinate.
It allows ultraviolet divergent contributions to be clearly isolated
and extracted using any convenient regulator (which does not
constrain momenta in the one non-trivial direction).
As shown by our example of scalar QED, the method remains applicable
in gauge theories, where multiple fluctuating fields are unavoidably
coupled.
Generalizing the treatment to, for example, full electroweak
theory is straightforward.

{\em Note added}: After the completion of this work, we became aware
of Refs.~\cite {Baacke1,Baacke2,Surig,Baacke3},
in which methods quite similar to those
we describe are applied to the closely related problem of
fluctuation corrections in bubble nucleation.

\section*{Acknowledgment}

We would like to thank Anton Ryzhov for reading
the manuscript and making a number of valuable suggestions.
This work was supported, in part, by the U.S. Department
of Energy under Grant No.~DE-FG03-96ER40956.

\appendix
\section{ Functional determinants from the solutions of associated
          differential equations}

The ratio of one-dimensional determinants (\ref{det1})
follows from the result that, for any bounded 
functions $U^{(0)}(x_{\bot})$ and $U^{(1)}(x_{\bot})$,
\begin{equation}
\label{det1_a}
    \frac{ \det_{\bot} \left(
		-\partial_{\bot}^2+U^{(0)}(x_{\bot})-\lambda 
		\right) }{
		\det_{\bot} \left(
		   -\partial_{\bot}^2+U^{(1)}(x_{\bot})-\lambda
		   \right)}=
	  \frac{
       \psi(\lambda;L) }{ \varphi(\lambda;L) },
\end{equation}
where $\psi$ and $\varphi$ are the solutions to the associated
ordinary differential equations,
\begin{eqnarray}
\label{deq1_a}
 \left[-\partial_{\bot}^2+ U^{(0)}(x_{\bot})-\lambda
 \right] \psi(\lambda;x_{\bot}) & = & 0,\\
\label{deq2_a}
  \left[-\partial_{\bot}^2+U^{(1)}(x_{\bot})-\lambda
 \right] \varphi (\lambda;x_{\bot}) & = & 0,
\end{eqnarray}
with boundary conditions specifying vanishing initial values
\begin{equation}
\label{bc_1_a}
 \psi(\lambda;-L)=0; \qquad \varphi(\lambda;-L)=0,\\
\end{equation}
and unit initial slopes,
\begin{equation}
\label{bc_2_a}
  \psi(\lambda;-L)'=1; \qquad  \varphi(\lambda;-L)'=1.
\end{equation}
Here primes denote $\partial/\partial x_{\bot}$,
and the functional determinants in (\ref{det1_a}) are
defined on the space ${\cal H}_L$
of functions vanishing at $x_{\bot}= \pm L$.

To prove the result (\ref{det1_a}), one need only notice
that both sides of this equation are meromorphic
functions of $\lambda$ with the same poles
and zeroes, and that both sides approach one
as $| \lambda | \rightarrow \infty$ in any direction
away from the positive real axis. Consequently
they must be identical.
This proof is  due to Coleman (see Ref. \cite{Coleman}).
It may be easily generalized to the case 
of higher-dimensional operators.
Let $W^{(a)}=\| W^{(a)}_{ij} \|$, $a=0,1$,
be two self-adjoint second order differential
operators acting on $\bigoplus_{i=1..n} {\cal H}_L$
of the form $ W^{(a)}_{ij}=-\partial_{\bot}^2+U_{ij}^{(a)}(x_{\bot})$.
Then by the same argument sketched above,
\begin{equation}
\label{det2_a}
  \frac{ \det_{\bot} (W^{(1)}-\lambda I) }{
		\det_{\bot} (W^{(0)}-\lambda I) }=
  \frac{\sum_{i_1...i_n}\epsilon_{i_1...i_n} \psi^{(1)}_{i_1}(\lambda;L) 
           ... \psi^{(n)}_{i_n}(\lambda;L) }{\sum_{j_1...j_n}
       \epsilon_{j_1...j_n} \varphi^{(1)}_{j_1}(\lambda;L)
           ... \varphi^{(n)}_{j_n}(\lambda;L) },
\end{equation}
where $I$ is the unit matrix, and
$\psi$'s and $\varphi$'s are the solutions of
the associated system of ODEs,
\begin{eqnarray}
\label{deq21_a}
  \sum_j (W^{(1)}_{ij}-\lambda \delta_{ij})
            \psi^{(k)}_j(\lambda; x_{\bot})&=&0, \\
  \sum_j (W^{(0)}_{ij}-\lambda \delta_{ij})
             \varphi^{(k)}_j(\lambda; x_{\bot})&=&0,
\end{eqnarray}
with the initial conditions
\begin{eqnarray}
\label{bc_21_a}
  \psi^{(k)}_i(\lambda;-L)= 0; \qquad \psi^{(k)}_i(\lambda;-L)'=\delta_{ik},\\
  \varphi^{(k)}_i(\lambda;-L)= 0; \qquad \varphi^{(k)}_i(\lambda;-L)'=\delta_{ik}.\\ \nonumber
\end{eqnarray}
The solutions 
\{$\psi^{(1)}(x_{\bot}),...,\psi^{(n)}(x_{\bot})$\}
are linearly independent, and therefore the numerator
in the right-hand side of (\ref{det2_a}) vanishes if and only if
there is a non-trivial solution of 
$W^{(1)} \psi(\lambda; x_{\bot})=\lambda \psi(\lambda; x_{\bot})$
which vanishes at $x_{\bot} = \pm L$.
In other words, the numerators of both the left and right
hand sides of Eq. (\ref{det2_a}) vanish whenever $\lambda$
is an eigenvalue of the linear operator $W^{(1)}$ (defined
on the interval $[-L,L]$ with Dirichlet boundary conditions).
The same argument applies to the denominators on either side.
One may also show that both sides approach unity as 
$|\lambda| \rightarrow \infty$ in any direction away
from the positive real axis, and therefore conclude that
both sides must be equal.

The specific relation (\ref{det_ntwo}) immediately
follows as the special case of $n=2$,
$W^{(1)}=\tilde{\Delta}_{A\phi}$,
$W^{(0)}=\tilde{\Delta}^{(0)}_{A\phi}$,
and $\lambda=0$.

\end{document}